\documentclass[12pt,a4paper]{article}
\usepackage{jheppub}
\usepackage[FIGTOPCAP]{subfigure}
\usepackage{amsmath}
\usepackage{amsfonts}
\usepackage{hyperref}
\usepackage{slashed}
\usepackage{amssymb}
\usepackage{tensor}
\usepackage{booktabs}
\usepackage{amsmath}
\usepackage{amssymb}
\usepackage{amsthm}
\usepackage{psfrag}
\usepackage{graphicx}
\usepackage{color}
\usepackage{bbm}
\usepackage{color}

\newcommand{\braket}[1]{\langle{#1}\rangle}

\newcommand{\exclude}[1]{}

\newcommand{\beq}{\begin{equation}}
\newcommand{\eeq}{\end{equation}}
\newcommand{\bea}{\begin{eqnarray}}
\newcommand{\eea}{\end{eqnarray}}
\long\def\/*#1*/{}

\newcommand{\p}{\partial}

\newcommand{\OO}{{\cal O}}

\newcommand{\PPi}{{\cal E}}

\newcommand{\junk}[1]{}

\title{\center{On Effective Holographic Mott Insulators }}

\author[1]{Matteo Baggioli\note{mbaggioli@ifae.es},}
\author[2]{Oriol Pujol\`as\note{pujolas@ifae.es}}

\affiliation[]{Institut de F\'{i}sica d'Altes Energies (IFAE), Universitat Aut\`{o}noma de Barcelona,\\ The Barcelona Institute of
Science and Technology, Campus UAB,\\ 08193 Bellaterra (Barcelona), Spain}

\abstract{We present a class of holographic models that behave effectively as prototypes of Mott insulators -- materials where electron-electron interactions dominate transport phenomena. 
The main ingredient in the gravity dual is that the gauge-field dynamics contains self-interactions by way of a particular type of non-linear electrodynamics. 
The electrical response in these models exhibits typical features of Mott-like states:  
i)  the low-temperature DC conductivity is unboundedly low; 
ii) metal-insulator transitions appear by varying various parameters;
iii) for large enough self-interaction strength, the conductivity can even decrease with increasing doping (density of carriers) -- which appears as a sharp manifestation of `traffic-jam'-like behaviour;  
iv) the insulating state becomes very unstable towards superconductivity at large enough doping. We exhibit some of the properties of the resulting insulator-superconductor transition, which is sensitive to the amount of disorder in a specific way.
These models imply a clear and generic correlation between Mott behaviour and significant effects in the nonlinear electrical response. We compute the nonlinear current-voltage curve in our model and find that indeed at large voltage the conductivity is largely reduced.

}

\begin{document}

\maketitle

\section{Mottivation}\label{sec1}

Strongly correlated materials are interesting because interactions play a very significant role and therefore they are not easy to describe. Examples of these materials include basically all known high-temperature superconductors, strange metals, correlated insulators, etc. There is a vast literature on these materials and on the various techniques to describe them, see e.g. \cite{Mott2,Mott3,Mott4,DMFT,Lee:2006zzc} for reviews. 
It is widely accepted that one can distinguish 3 different mechanisms  that can be responsible for the nontrivial (electrical) response: 
electron-phonon (e-ph), electron-disorder (e-dis), and electron-electron (e-e) interactions. Usually, Mott insulators \cite{Mott1,Mott6,Mott5} refer to the materials that are dominated by the latter: charge-carrier self-interactions. 
Of course, in real materials a combination of them all might be relevant, but it is important to distinguish and understand them separately. The heuristic picture that summarizes the Mott behaviour (sometimes referred to as Mottness) is that of an {\em electronic traffic jam}: strong enough e-e interactions should, of course, prevent the available mobile charge carriers to efficiently transport charge. The purpose of this article is to use the bottom-up version of the gauge-gravity duality as a low-energy effective model for this type of materials.

There has been a remarkable progress recently in understanding how the gauge-gravity duality  methods can be adapted for condensed matter problems. What the holographic models accomplish to do is to give explicit and nontrivial ({\em i.e.}, interacting) yet tractable field theories that include various operators and which can be used to model the limit of  strong-correlations and criticality. For instance, it is very well understood how to construct gravitational models in 3+1 dimensions that behave as 2+1 CFTs with a charge current $J_\mu$ and a stress tensor $T_{\mu\nu}$ operators: these are simply Einstein-Maxwell theories in asymptotically AdS spacetimes.   We also know how to introduce the breaking of translations, which is certainly a crucial ingredient for the condensed matter applications since this includes both phonons and disorder. There are various ways to introduce translation breaking sector, but one of the most convenient ones is through a set of marginal operators $O^I$ \cite{Andrade:2013gsa,Baggioli:2014roa,Taylor:2014tka,Andrade:2016tbr,Andrade:2015hpa}. In the gravity picture, these models reduce to Einstein-Maxwell-Stueckelberg theories, or what is the same, to a certain class of Maxwell - Massive Gravity theories \cite{Rubakov:2008nh,Vegh:2013sk,Davison:2013jba,Alberte:2015isw,Baggioli:2015gsa}.

Two basic messages from these recent developments are:
\begin{enumerate}
\item[i)] the bottom-up version of the gauge-gravity duality provides an effective description at low energies. This should be taken strictly in the sense of Effective Field Theories (EFTs) that are formulated directly in terms of low-energy degrees of freedom (the $T_{\mu\nu}$, $J_\mu$ and $\OO^I$ operators and the excitations contained therein), which has the advantage that it represents an efficient re-summation of all the non-trivial interactions. Therefore, there is absolutely no reference to the microscopic structure of the material. Still, the  low energy observables such as the transport parameters (and the various constraints among them) are neatly accessible in a controlled way. A big difference with respect to standard EFTs is that instead of having an energy-gap in the mass spectrum of excitations, one has a gap in the spectrum of scaling dimensions of the various operators. This is the key ingredient that allows for a well-defined notion of effective conformal theory which, in turn, allows to study and model strongly coupled systems with critical or scaling behavior. 

\item[ii)] interactions amongst the various sectors that participate in the CFT have a clear counterpart in the gravity side where the dynamics unfolds in a rather standard, local and classical field theory formulation. For instance,  the e-ph and e-dis interactions stem from a various types of interactions between the Maxwell and Stueckelberg sectors \cite{Baggioli:2014roa,Baggioli:2016oqk,Gouteraux:2016wxj}. 
\end{enumerate}

In this article we will focus on modeling materials that are dominated by e-e interactions with the same kind of effective holographic methods. Since the electron or charge sector is encoded in the charge current operator $J_\mu$, and this is incarnated holographically in the Maxwell gauge field $A_\mu$, introducing self-interactions in the charge sector clearly requires to introduce self-interactions for the $U(1)$ gauge field $A_\mu$. The only way to introduce such self-interactions while preserving gauge-invariance is that the  the field strength $F_{\mu\nu}=\nabla_\mu A_\nu-\nabla_\nu A_\mu$ appears in the Lagrangian beyond quadratic level. In other words, that the gauge-field Lagrangian in the gravity side is what is usually called Nonlinear Electrodynamics (NED).

The main point that we are going to illustrate is that the effective holographic models with a Nonlinear Electrodynamics (NED) of certain type: i) provide simple, tractable and consistent models that are naturally fit to model Mott-physics; and ii) the  phenomenology of these models matches well with this interpretation. For instance, one can easily realize a metal-insulator transition (MIT) that is clearly driven by ``e-e interactions'', that is, by the nonlinear structure of the Maxwell/charge sector. In the examples below, there is no other dynamical ingredient (such as a non-trivial renormalization group flow at $T=0$, or a significant amount of disorder) that plays any role in the MIT, therefore we find that these MITs are clearly driven by the e-e interactions.  

We shall then consider non-linear extensions of the Maxwell theory,
that is to include nonlinear terms in the gauge field action such as
$$
{F_{\mu\nu}F^{\mu\nu}}+(F_{\mu\nu}F^{\mu\nu})^2+(\tilde F_{\mu\nu}F^{\mu\nu})^2+...
$$
It is of course not necessary for our purposes to consider the most general NED theory, but since we are certainly interested in the strong field (nonlinear) regime, then it is more convenient to assume from the start that the Lagrangian depends on the field strength  invariants through a generic functional form.   
For the sake of simplicity, it will suffice to consider the class of NED models with a Lagrangian of the form:
\begin{equation}\label{choice}
K\big({F_{\mu\nu}F^{\mu\nu}}\big)
\end{equation}
with $K$ a generic function. Let us now remark a few points. 

First of all, considering functions of $F_{\mu\nu}F^{\mu\nu}$ (and $\tilde F_{\mu\nu}F^{\mu\nu}$) is perfectly compatible with the EFT logic: it can be understood as a truncation of the action to all the nonlinear  terms that are of first order in derivatives, which is relevant in the regime where the fields are strong and with small gradients. This is exactly the kind of limit where the  DBI action is a good approximation for certain higher dimensional extended objects with localized gauge fields. Note also that, as the DBI case makes manifest, in some cases the full functional form of $K$ can be protected by symmetries.  

Second, including higher powers of $F_{\mu\nu}F^{\mu\nu}$ {\em is} relevant even for the linear electrical response because we are interested in black brane solutions with a nonzero charge density (which maps to the density of mobile charge carriers). In these solutions, $F_{\mu\nu}$ has a nonzero background value and all terms in the infinite series \eqref{choice} can therefore contribute to the linear conductivity. Indeed, in Reissner-Nordstr\"om black brane solutions $F^2$ grows large close to the horizon, so one can foresee that that gauge field nonlinearities can considerably affect the near-horizon region (the IR properties of the charged CFT plasma such as the DC conductivity).

Third, the choice of nonlinear model \eqref{choice} is by far not the most general one, but we stick to it just as a minimal model that includes self-interactions of the Maxwell/charge sector.  
We exclude other high dimension operators in the bulk such as  $\tilde F_{\mu\nu}F^{\mu\nu}$ and powers thereof just  for simplicity; and $(\nabla F)^2$ and powers thereof because they increase the order of the equations of motion, so they have to be treated as perturbations.  Similarly, we leave out other operators involving the charge and translation breaking sector because these would correspond to more complicated cross-interactions between charge carriers to  disorder and/or phonons.

All in all, we just take \eqref{choice} as  a minimal, consistent and effective model that allows us to explore the effect of these charge sector nonlinearities by themselves in the transport properties. As already emphasized, we stick to the simplest (and most challenging) class of models that produce {\em no mass gap} (down to $T=0$) and which still contain a {\em nonzero density of charge carriers}. In the holographic models, we identify the density of charge carriers simply as the charge density of the charged black brane solutions, $\rho$, since the `elements' of the black brane horizon certainly carry charge and are mobile to some extent. 
Since in actual materials the density of charge carriers can be controlled externally by electron- or hole- `doping', one can further identify the BB charge density $\rho$ as the  {\em electron/hole doping}.

\paragraph{Summary of results:}
Next we summarize our  main results, which we divide in 4 groups.\\[-2mm]

1) Spotting Mott insulators in {\em holographic nonlinear electrodynamics}:\\[2mm]
- We study general Nonlinear Electrodynamics (NED) models of the type \eqref{choice} with arbitrary kinetic function $K(z)$ embedded in the holographic setup. We find the NED charged, asymptotically AdS, planar Black Branes solutions to these models, the analogues of the `Reissner-Nordstr\"om' solutions for NED theories. We find the consistency constraints on the choice of $K(z)$ that ensure that the model is free from instabilities or other pathologies. We then study their linear and nonlinear electric response for the models that pass the consistency constraints.\\[-4mm]

- We identify  an especially interesting and simple class of NED to study defined by a 1-parameter family of `benchmark' DBI-like models, endowed with a parameter, $\Theta$, that encodes the strength of the nonlinearities. For the standard DBI case one has $\Theta<0$, which exhibits metallic properties. We extend the analysis to the continuation to $\Theta>0$ (which is ghost-free version of a DBI model with negative  tension).  We dub this case {\em iDBI} since it displays insulating properties. \\[-4mm]

-  Our benchmark holographic  models contains then basically 4 parameters (aside from the AdS length $\ell$): i) the BB charge density  $\rho$; ii) the graviton 'mass' $m$; iii) the gauge self-coupling $\Theta$ and iv) the gauge-metric coupling $q$ (that controls how much backreaction the gauge sector gives on the metric). We find that a consistent condensed matter re-interpretation of these parameters is as follows. $\rho$ (with dimensions of charge density) corresponds to the net mobile charge-carrier density. In some material this can be externally dialed by means of electron/hole doping, so in practice one can identify $\rho$ with the electron/hole doping (of course up to the offset defined by the carrier density at zero electron/hole doping). $m^2$ can be identified with the disorder strength, concretely as a (homogeneous) density of electrically neutral impurities (which do not interact directly with carriers \cite{Baggioli:2016oqk}, such as for instance by not providing electrons/holes). $\Theta$ is dimensionless and characterizes the strength of unscreened e-e interactions for the low-energy charge carriers. {Note that in our models both  signs for $\Theta$ are in principle allowed. Different signs for $\Theta$ lead to opposite kind of self-interaction that could be labelled as `attractive' or `repulsive'. It is quite indirect how this effective interaction relates to the interaction between  individual carriers (electrons and/or holes). In real materials, e-e interactions are repulsive but electron-hole interactions are attractive, and so both signs could be relevant to model different types of material/carrier composition. For Mott-like insulators (dominated by one-sign carriers), however, one would expects that only one sign of $\Theta$ will manage to mimic the e-e (repulsive) interactions. Indeed, we find that the models reproduce Mott-like behaviour only for $\Theta>0$.\color{black}}
Finally $(q\ell)^2$ from the CM perspective  controls the mixing between the current density and the momentum density as well as perhaps the fraction of the energy density  that is stored in e-e interactions. This last identification is admittedly more speculative but also less central to our work.\\[-2mm]

- The iDBI model at the level of the background already presents quite interesting properties. For values of the self-coupling $\Theta$ larger than a certain threshold $\Theta_1$  ($\Theta_1 = (q\ell)^2/3$), the background solutions admit only up to a maximum charge density $\rho_\star$.\footnote{For $\rho>\rho_\star$ the solutions have a naked singularity. We shall not address here whether or not these naked singularities are admissible and can be given a physical interpretation along the lines of \cite{Gubser:2000nd}. Rather, we will focus only on the cases where the singularity is hidden by a horizon, that is, to $\rho<\rho_\star$.} 
The value of $\rho_\star$ depends on the disorder strength (density of impurities) and the self-interaction strength. Indeed, in the simplest models below we will find (see Eq. \eqref{rho-bound}) a linear relation between the maximal carrier density $\rho_\star$ and the disorder strength (density of impurities, $m^2$)
\beq\label{crit-dop-dis}
\rho_\star = F(\Theta, q\ell ) \;m^2
\eeq
The fact that for close to this upper bound the geometry and gauge-field configurations become more and more singular will imply that the system becomes more and more unstable.
In particular, for $\rho<\rho_\star$ but close to the maximum $\rho_\star$ the solutions are expected to develop a  superconducting instability (see below).

2) Linear electrical response: \\[-8mm]
\begin{itemize}
\item[-] We obtain the DC conductivity $\sigma_{DC}$ in these models in terms of horizon data and the AC conductivity $\sigma(\omega)$ numerically.\\[-8mm]
\item[-] The consistent models of type \eqref{choice} can be split in two classes, metallic and insulating, according to whether $\sigma_{DC}$ at $T=0$ is enhanced or reduced with respect to the Maxwell (linear) theory. In  metallic/insulating models, the slope of the the kinetic function $K$ at large argument is bigger/smaller than for the Maxwell case respectively. \\[-8mm]
\item[-] We construct insulators that have very low $\sigma$ and still finite $\rho$ and no confining/hyperscaling IR dynamics. We find that this is enough to illustrate that it is possible to implement insulating phases even in the limit of conformal behaviour and which is clearly driven by the self-interactions of the charge sector, at least in the context of Effective Holographic models (or bottom-up holography).\\[-8mm]
\item[-] We construct a simple transition between the metal and insulator regimes that occurs by dialing  the  parameter that characterizes the self-coupling. Therefore, we find that this is a sharp example of a MIT that is driven by charge sector self interactions.\\[-8mm]
\item[-] These results also exhibit that there is no lower bound on the electrical conductivity in effective holographic models, and that charge-sector self-interactions can be a mechanism to violate the bound that is present in the restricted class of Maxwell (linear) models \cite{Grozdanov:2015qia}.\\[-8mm]
\item[-] The iDBI cases exhibit another rather interesting phenomenon. As mentioned before, above a first threshold, $\Theta>\Theta_1=(q\ell)^2/3$, an upper bound on the  carrier density appears, $\rho<\rho_\star$. By further increasing $\Theta$ (and for $\rho<\rho_\star$) a second threshold $\Theta_2$ appears (we find  $\Theta_2 = 2 \Theta_1$) above which  the   DC conductivity even becomes a {\em decreasing} function of $\rho$ (at $T=0$). We find this property to be a quite sharp manifestation of  Mott-like (electronic traffic jam) behaviour, that further confirms our interpretation. \\
\end{itemize}

\begin{figure}
\centering
\includegraphics[width=10cm]{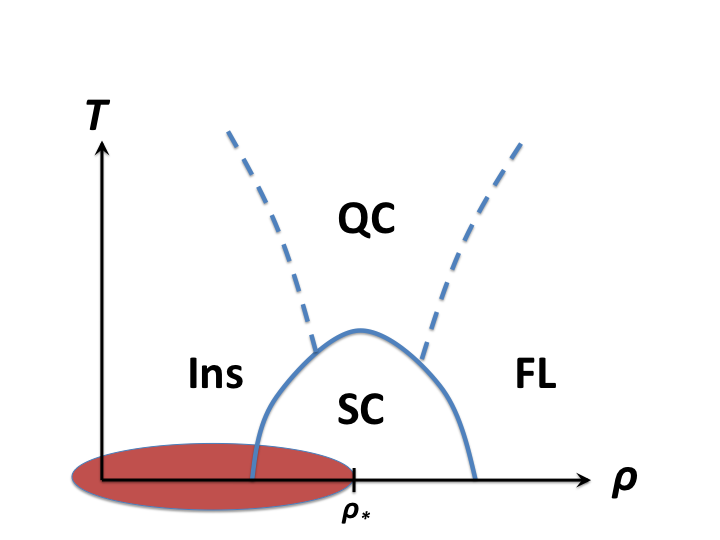}
\caption{Sketch of the phase diagram of a strongly correlated material. {\em Ins} stands for insulating phase, {\em SC} for superconducting phase, {\em FL} for the Fermi Liquid Metallic phase and {\em QC} for the Quantum Critical region which has usually a Strange Metal behaviour. The red-shaded area refers to the  Insulator-Superconductor transition that takes place generically in the presently studied iDBI models at low temperatures.}
\label{SCsketch}
\end{figure}

3) Superconductivity:\\[-5mm]
\begin{itemize}
\item[-] Above the first self-interaction threshold $\Theta_1$, an upper bound on the carrier density appears $\rho<\rho_\star$. Near the upper bound, the holographic gauge-field  configuration becomes more and more singular and one can expect SC instability to appear. We study this by allowing a charged scalar condensate to be dynamical and analyzing its possible instabilities at $T=0$. We find indeed that a SC transition is generically enhanced in the iDBI models, and that SC appears before the upper bound, at a certain $\rho_{SC}<\rho_\star$ in the iDBI models. This happens quite robustly and independently of the scalar condensate properties (such as the scaling dimension of the scalar operator in the CFT language).
\item[-] This fits well into the qualitative picture of the Hight Tc SC phase diagram, and with the identification that the bulk charge density $\rho$ plays the role of electron/hole doping. This suggests also that the  SC `dome' in the phase diagram (see Fig.~\ref{SCsketch}) in the electron/hole doping axis should be placed around $\rho_\star$.   
\item[-] This suggests that of the `critical doping' $\rho_{SC}$ where we have the SC quantum phase transition has a fixed dependence on disorder. Since $\rho_{SC}$ appears necessarily because of the upper bound, one expects that the $\rho_{SC}$ depends on disorder in a similar way as $\rho_{\star}$, that is like in \eqref{crit-dop-dis} -- the critical electron/hole doping should increase with the disorder strength. 
\end{itemize}

4) Nonlinear electric response:\\
It is clear that the main robust and generic feature of NED charged AdS black branes is that they enjoy a large nonlinear electric response. We illustrate this by computing:
\begin{itemize}
\item[-]  The nonlinear charge-density vs chemical potential curve $\rho-\mu$. The insulating iDBI models the curve is always below the linear Maxwell and metallic DBI ones. \\[-8mm]
\item[-] The nonlinear DC conductivity, that is the nonlinear current vs voltage $I$-$V$ curve (which is equivalent to the $J-E$ curve) in the `probe' approximation. Again, in the insulating iDBI models the curve is always below the linear Maxwell and metallic DBI ones. The probe analysis suggests that the material can withstand a up to a finite maximum voltage.\\[-8mm]
\color{black}
\end{itemize}

~\\
Previous literature on holographic models dual to Mott insulators without introducing Nonlinear Electrodynamics in the bulk include \cite{Edalati:2010ww,Wu:2012fk,Edalati:2010ge,Ling:2014bda,Fujita:2014mqa,Ling:2015epa,Nishioka:2009zj,Kiritsis:2015oxa,Donos:2012js}. The main difference between the present work and the latter is that in our model the insulating behaviour\footnote{See \cite{Rangamani:2015hka,Donos:2014uba,Donos:2013eha} for other holographic realizations of Insulators and MITs driven via other mechanisms.} is clearly only driven by charge-carrier self interactions even in a situation where there is no hard gap and that the analysis is performed in a full backreacted fashion (which is not available, yet, for the probe fermions models for example).\\
In \cite{Myers:2010pk} non linear corrections to the Maxwell term in the bulk were partially discussed (though not in connection to Mott physics) and some effects on the linear conductivity similar to the ones we see were found.\\[0.3cm]
\color{black}
The rest of this article is organized as follows. In Section \ref{sec2} we define the family of models and background black brane solutions that we will study. We will discuss the constraints which  the kinetic function $K(z)$ must satisfy and will define a benchmark model that satisfies them and allows to illustrate the main features in function of some simple parameters.
In Section \ref{sec3} we present the main phenomenological features of these models. These  include i) the presence of MIT transitions that  result from the charge-sector self interactions, ii) the absence of any lower bounds on the conductivity \footnote{See \cite{Grozdanov:2015qia,Grozdanov:2015djs,Hartnoll:2016tri,Burikham:2016roo,Alberte:2016xja,Amoretti:2014ola,Hartnoll:2014lpa,Ge:2015fmu,Ikeda:2016rqh,Blake:2016wvh,Blake:2016sud} for interesting discussions related to the existence or not of universal bounds in the context of momentum dissipating holography.}, and iii) the presence of significant nonlinear response.
In Section \ref{SCsection} we discuss the SC instability and the presence of insulator-SC transitions. 
\color{black} 
In Section \ref{probesection} we discuss a simplified version of the model, namely the probe limit where the fluctuations of the background metric are completely neglected but still some of the effects of the charge sector self-interactions persist.\\
\color{black}
We conclude with some discussion in Section \ref{discussect} and with technical details in the Appendices \ref{app1}, \ref{app2}, \ref{NLgeneral}.

\section{Modeling e-e interactions with holography}\label{sec2}

We consider a model with 3 sectors: the metric $g_{\mu\nu}$ of a 3+1 dimensional spacetime, a $U(1)$ gauge field $A_\mu$ and a the Stueckelberg fields $\phi^I$ with $I=1,2$, which break translational invariance. As discussed above, we want to discuss asymptotically AdS solutions and a model where the gauge sector enjoys non-linear Lagrangian. A representative model that contains self-interactions in the gauge and Stueckelberg sectors separately has an action of the form,
\begin{equation}
\mathcal{S}\,=\frac{M_P^2}{2}\,\int d^4x\,\sqrt{-g}\,\left[R-2\,\Lambda+\,q^2\,K\left(-\frac{F^2}{2}\right)-2 \,m^2 \,V(X)\right].\label{action}
\end{equation}
Here, $X=g^{\mu\nu}\p_\mu\phi^I\p_\nu\phi^I$, $F^2=F_{\mu\nu}F^{\mu\nu}$, $F_{\mu\nu}=\partial_\mu A_\nu-\partial_\nu A_\mu$, 
$\Lambda=-3/\ell^2$ with  $\ell$ is the AdS radius and $M_P^2$ the bulk Planck mass. 
Both the charge and Stueckelberg sectors contain two mass scales: $q$ or $m$, which measure the backreaction of these sectors to the metric. Additionally they have an intrinsic scale that suppresses the nonlinear terms implicit in $K(F^2)$ and $V(X)$, which 
is implicit  in the functional form of $K$ and $V$\footnote{We find convenient to use a normalization for $A_\mu$ such that it has dimensions of length. This relates to the canonical normalization as $A_{\rm can}^\mu$ as $A^\mu= (\ell/M_P) A_{\rm can}^\mu$. In terms of $A_{\rm can}^\mu$, the gauge field Lagrangian  reads $ M_P^2 q^2 K( -F_{\rm can}^2 / M_{gauge}^4)$, with $M_{gauge}$ the gauge self-interaction scale. In the benchmark model below \eqref{powermodel} , $K(z)$ is really a function of $\Theta z$ with $\Theta$ a constant that parameterizes the gauge field  self-coupling. The scale of the gauge nonlinearities is then given by $M_{gauge}^4= M_P^2 \ell^{-2} /\Theta$. 
%
}.
We discuss the constraints on the possible forms of $K(z)$ below. As mentioned above, this is not the most general form of the action for neither the gauge nor the Stueckelberg sectors. Still it is enough to illustrate our points.  

Let us emphasize that there is no problem in assuming non-canonical kinetic terms for $\Phi$ or $A_\mu$ from the Effective Field Theory point of view in the bulk, precisely because we allow for a general kinetic function. The full form of $K(z)$ can be protected by some symmetry, as is the case for instance for Dirac-Born-Infeld (DBI) models. In that case the nonlinear scale is set by the brane tension which can be parametrically smaller scale than the Planck mass. Indeed, for the models \eqref{powermodel} with $\Theta\gg1$, then one can safely study the strong field configurations with  $M_P^4 \gg F_{\rm can}^2 \gg M^4_{gauge}$. In addition, in the $q \ell \ll1$ limit, the gravitational backreaction can be neglected -- this is the `probe' limit.

Let us first exhibit the form of the asymptotically Anti de-Sitter planar black brane solutions of this model. Using coordinates $\{t,x,y,u\}$ (with $u$ parameterizing the `holographic' direction) they can be obtained by the following ansatz,
\begin{align}
&ds^2 = \frac{\ell^2}{u^2}\left[-f(u)dt^2+\frac{1}{f(u)}du^2+dx^2+dy^2\right]\nonumber\\
&\phi^I=\alpha \,\delta^I_i x^i\nonumber\\
&A_t(u)=\ell^2\int_{u}^{u_h} \frac{\PPi(t)}{t^2}\,dt
\label{ansatz}
\end{align}
where $\PPi$ plays the role of the local electric field in the $u$ direction, $(-F_{\mu\nu}F^{\mu\nu}/2)^{1/2}$, on the solution. 
The Einstein equations and the Maxwell equations then reduce to (setting $\ell=1$):
\begin{align}
&4\,\Lambda+12\,f(u)+4\,m^2\,V(u^2 \alpha^2)-2\,q^2\,K(\PPi^2)-4\,u\,f'(u)+4\,q^2\,\PPi^2\,K'(\PPi^2)=0\nonumber\\
& K'(\PPi(u)^2)\, \PPi(u)=\rho\, u^2
\label{eoms}
\end{align}
The Einstein equation then allows to express the emblackening factor $f(u)$ of the metric to take the closed form,
\begin{equation}
f(u)=u^3 \int_{u}^{u_h}\left(-\frac{\Lambda}{y^4}-\frac{m^2\,V(y^2\alpha^2)}{y^4}+\frac{q^2\,K(\PPi(y)^2)}{2\,y^4}-\frac{q^2\,y^2\,\rho\,\PPi(y)}{y^4}\right)\,dy
\label{BlackFactor}
\end{equation}
which for any given choice of $K$ and $V$ can be immediately worked out. 
The Maxwell equation of course becomes non linear, and in the linear case $K(z)=z$ one obtains the usual, $\PPi(u)=\rho\,u^2$ and $A_t=\mu-\rho\, u$. For general choices of $K$ working out $\PPi(u)$ is obtained only implicitly (and one may need to perform numerically the integral to obtain $A_t$). In any case, we will restrict ourselves to the simplest cases where one can obtain analytic expressions.
Lastly, note that the choice $K(F^2)$ breaks the S-Duality of the Maxwell sector in the bulk (corresponding to  particle-vortex duality of the dual CFT as in \cite{Myers:2010pk}) but it in principle there might be nonlinear choices of the Lagrangian that do not break it (see {\em e.g.} \cite{Gibbons:1995cv,Gibbons:1995ap}).

\subsection{Consistency constraints}

Playing with a non linear charge sector we need to be careful and ensure that the system is consistent and healthy. In order to do so we first of all impose that:
\begin{equation}
K(0)\,=\,0\,,\,\,\,\,\,\,K'(0)\,=\,1\,.
\end{equation}
which essentially means that for small field strengths we recover the standard Maxwell theory which is encoded in a linear $K$ function $K(z)=z$.

Non linearities in the charged sector in the form of eq.\eqref{choice} may additionally give rise to issues of consistency in the form of ghosty perturbations and/or gradient instabilities. We discuss in details the possible problems appearing because of that choice in appendix \ref{app1}. 
For the sake of clarity, we summarize here the results of the appendix \ref{app1}.
All in all, the minimal constraints for the model \eqref{choice} read:
\begin{align}
K(0)\,=\,0\,,\,\,\,\,\,\,K'(0)\,=\,1\,,\,\,\,\,\,\, K'(z)\,>\,0\,, \,\,\,\,\,\,\left(\sqrt{z}\,K'(z)\right)'\,>\,0\,
\label{textconstr}
\end{align}
and will be fully satisfied by the models considered throughout this paper.
The last condition in \eqref{textconstr} arises from the absence of gradient instabilities in the scalar modes of the gauge-field perturbations. Since the analysis of these modes is rather lengthy we restricted it to the decoupling limit, that is, to the cases where the mixing with scalar modes in the metric can be neglected. This corresponds to $q\ell\ll1$, but one does not  expect major differences to appear at larger values of $q$~. These constraints are less simple to express in full generality but translate into an upper bound on the strength of the gauge-field nonlinearities in the sub-class of models where $K'(z)<1$, which will affect the models of insulating type below.

Note that the form of  $K(z)$ that is on the verge of violating the last condition is $\sqrt{z}$. An interesting subclass of models that is close to this behaviour is the DBI-like choice $\sim\sqrt{1+z}$, where one gets close to this behaviour at large $z$.
We will indeed concentrate on this kind of models, which is where more dramatic effects arise and some analytic results can be obtained.

Along with the previous constraints we remind the ones for the non linear translational symmetry breaking sector which were extracted in \cite{Baggioli:2014roa} and which take the following structure:
\begin{align*}
V(0)\,=\,0 \qquad V'(0)\,=\,1 \qquad \,V'(X)\,>0\,,\qquad\,V'(X)\,+\,{X\,V''(X)}\,>0\,.
\end{align*}

\subsection{Benchmark model}
As mentioned above, we consider generic non-linear electrodynamics (NED) models where we replace the usual Maxwell term by
\begin{equation}
-\frac{F_{\mu\nu}F^{\mu\nu}}{2}\,\,\,\,\longrightarrow\,\,\,\,K \left(-\frac{F_{\mu\nu}F^{\mu\nu}}{2}\right)
\end{equation}
where for the sake of simplicity we assume dependence on $F_{\mu\nu}F^{\mu\nu}$ only.

One simple class of models that interpolates between $K=z$ at small $z$ and a generic power at large $z$ takes the following form
\begin{equation}
K(z)\,=\,\frac{1}{\Theta\,p}\left(1+\Theta\,z\right)^p-\frac{1}{\Theta\,p}
\label{powermodel}
\end{equation}
In the limit $\Theta\rightarrow 0$ the model \eqref{powermodel} reduces to the Maxwell case (for every choice of p), while increasing $\Theta$ in the positive and negative directions we depart consistently from it.
The choice \eqref{powermodel} represents a higher order deformation of the usual Maxwell action:
\begin{equation}
K\left({-F^2\over2}\right) =-\frac{F^2}{2}+\,\Theta\,\frac{(p -1) }{8 \,}F^4-\,\Theta^2\,\frac{(p -2) (p -1) }{48}\,F^6+\,\dots
\label{Maxpert}
\end{equation}
From this expression is clear that $\Theta$ encodes  the non-linearity scale (the `critical field' where nonlinearities are important being $\propto\sqrt{|\Theta|}$). Since this choice complies with $K'(0)=1$ and $F^2\to0$ in the UV region, then in all of our charged  black brane solutions the nonlinearities are going to be noticeable only close to the horizon. 

The properties of these models differ a lot depending on the sign of $p-1$ and of $\Theta$. For the rest of the paper we will focus our attention on the DBI-like models (with power $p=1/2$) but we will allow $\Theta$ both positive and negative. Therefore, we shall refer to these models as 
\begin{align}
{\rm DBI}\,\,\,\,\,\,&\longrightarrow\,\,\,\,\,p\,=\,1/2,\;\;\Theta<0\,\,\,\,\nonumber\\
{\rm iDBI}\,\,\,\,&\longrightarrow\,\,\,\,\,p\,=\,1/2,\;\;\Theta>0.\label{models}
\end{align}
and `iDBI' is meant to remind that for $\Theta>0$ we will obtain {\em insulating} behavior.
For both signs of $\Theta$,  the constraints \eqref{textconstr} are satisfied\footnote{One can see the attractive/repulsive nature of these self interactions for $\Theta>0$ or $\Theta<0$ respectively from Eq.~\eqref{Maxpert}. At low fields, the scattering of electromagnetic waves is dominated by this the quartic term, which is proportional to $\Theta$. Clearly, the effect of the interaction (whether the waves tend to attract or to repel) is dictated by the sign of $\Theta$. Therefore only one sign can correctly match with the repulsive e-e interactions that we are trying to model. This is going to be $\Theta>0$, which corresponds to the `negative tension' iDBI case.
}.

Assuming the choice \eqref{models} the solution for the $\PPi$ function reads:
\begin{equation}\label{ppiu}
\PPi(u)\,=\,\frac{u^2\,\rho}{\sqrt{1-u^4\,\rho^2\,\Theta}}
\end{equation}
The solution satisfies $\PPi>0$ and $\PPi'>0$ and it is then safe from consistency issues and therefore healthy (see Appendix \ref{app1}).
%
%
For small $u$ (and large temperature $T$),  can be series expanded as:
\begin{equation}
\PPi(u)\,=\,\underbrace{\rho \, u^2}_{\text{Maxwell Term}}\,+\,\Theta\,\frac{\rho ^3\, u^6}{2}\,+\,O\left(u^7\right)\,\dots
\end{equation}
so it is clear that near the AdS boundary ($u=0$) the gauge field recovers the linear behaviour. Also, from this equation and \eqref{ansatz} it is clear that the integration constant $\rho$ still plays the role of the charge density.

Another feature  that is obvious from \eqref{ppiu} is that for $\Theta>0$ the solution can become complex-valued  at large enough $u$, that is sufficiently towards the IR. The condition on the gravitational backreaction will then be that the complex (singular) part of the solution is hidden by a horizon.

\begin{figure}
\centering
\includegraphics[width=8cm]{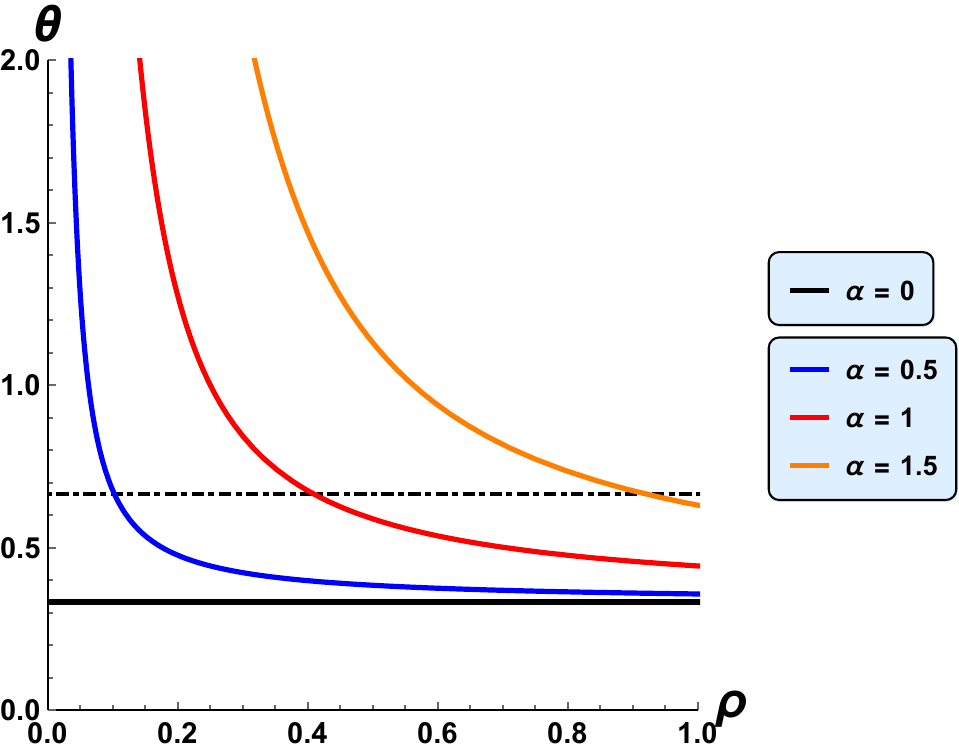}%
\qquad
\includegraphics[width=6.2cm]{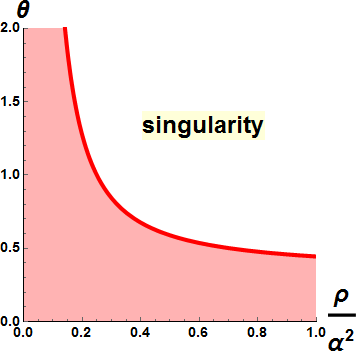}
\caption{\textbf{Left:} Phase diagram in the $\theta$-$\rho$ plane for different disorder strengths $\alpha$ with $q=1$. The different curves are $\rho_\star(\Theta)$ \eqref{rhostar} for various values of the disorder strength $\alpha$. The allowed regions (for the solution in order to have no naked singularities) are below the various curves. On the curves $\rho=\rho_\star$ and the extremal horizon coincides with the singularity. \textbf{Right:} Sketch of the situation rescaling $\rho$ by $\alpha^2$: the red shaded region defined the allowed parameter space.}
\label{probesigmatot1}
\end{figure}

Let us sketch the kind of constraints that one gets. For the models \eqref{models}, 
the canonical momentum that appears in   \eqref{eoms}, $K'(\PPi(u))\, \PPi(u)$, is a bounded quantity (in the iDBI case, $K' \, \PPi(u)\leq 1/\sqrt\Theta$). Since the equation of motion equates that quantity to $\rho\, u^2$ then the solution cannot extend arbitrarily into the infrared region. Therefore in the iDBI models the singularity is not  placed at $u\to\infty$ but instead at 
$$
u_\star\,=\,\rho^{-1/2}\, \Theta^{-1/4}
$$
where the solution formally becomes complex\footnote{Note that this situation arises only for models where $K(z)\sim z^{1/2}$ for large $z$.}. Note that  $u_\star$ is the place where $K'(u_\star)=0$, therefore one is at the verge of violating the ghost-freeness condition. One has to require then that this singularity is always hidden by the horizon, the most stringent constraint arising at extremality, $T=0$. The temperature of the black brane solutions in terms of the model (with $V(z)=z/2m^2$ so that now $\alpha$ plays the role of $m$) parameters and the horizon location reads:
\beq\label{T}
T\,=\,\frac{6-\alpha^2\, u_h^2-\frac{2\,q^2}{\Theta} \left(1-
   \sqrt{1-\rho ^2 \,u_h^4\,\Theta}\right)}{8\, \pi\,  u_h}\,,
\eeq
For $\Theta>0$, requiring that the singularity is hidden by a horizon gives $u_h<u_\star$, or
\beq\label{rho-bound}
\rho ^2 \,u_h^4\,\Theta <1
\eeq
That this happens all the way down to zero temperature (i.e. $u_h=u_0$, with $u_0$ the maximum value $u_h$ can take) then leads to 2 very distinct cases\footnote{This defines the first threshold mentioned above as
$\Theta_1 =  q^2 /3$.
}:
\begin{itemize}
\item $\Theta <  q^2 /3$: there is no problem to satisfy $u_h<u_\star$ at any $T$ and $\rho$. 
\item $\Theta>q^2/3$: $\rho $ needs to be below a certain maximum density, $\rho_{\star}$, in order that the singularity is hidden by a horizon. At $\rho_{\star}$, one has that $u_{\star}=u_0$, so\footnote{For $\alpha=0$, the condition  $u_h\leq u_{\star}$ all the way down to the extremal horizon translates into $\theta\leq q^2/3$.}
\beq\label{rhostar}
\boxed{\rho_{\star}=
{\alpha^2 \sqrt{\Theta} \over 2 (3 \,\Theta - (q\,\ell)^2)} }~.
\eeq
Note that for the $V(z)=z/2m^2$ choice $\alpha$ is playing exactly the role of the graviton mass, so \eqref{rhostar} corresponds precisely the perviously advertised relation between the charge-carrier density upper bound and the density of impurities, \eqref{crit-dop-dis}.  

\end{itemize}
All these features are summarized in Fig.\ref{probesigmatot1}. In the next Section we will get back to the characterization of the DC conductivity and the charge susceptibility in the different regions of parameter space.


Note that this has a finite limit when the backreaction from the charge sector is small $q\to0$. Therefore there is an upper bound on, $\rho<\alpha^2 / 6\sqrt\Theta$, that applies also in  the vector probe limit. Interestingly,  the maximal density of charge-carriers is basically set by the amount of disorder ($\alpha$) and the magnitude of the charge-carrier self-interactions ($\Theta$), once these overcome a certain threshold ($\Theta>q^2/3$). None of these features arise in the standard DBI case, in which the `sign' of the self-interactions is opposite.\\


The DBI case \eqref{models} corresponding to $\Theta<0$ has been  analyzed in depth in the literature and it corresponds to a metallic CFT \cite{Karch:2007pd}; the main novelty of our work refers to the iDBI (corresponding to $\Theta>0$) which gives rise to insulating CFTs.

As it is well known, the DBI case is protected by a symmetry, a non-linearly realized higher-dimensional Poincare group, see {\em e.g.} \cite{West:2000hr,Gliozzi:2011hj}. 
For this reason, the DBI action admits a geometrical interpretation as a brane with a localized gauge field that is embedded in higher dimensional space. The brane realizes the spontaneous breaking of the higher-dimensional Poincare group down to the 3+1 dimensional one.  The action then is fixed by this nonlinearly realized symmetry  to the familiar Nambu-Goto form $\sim (q^2/\Theta) {\rm Det}\Big({g_{\mu\nu}+\sqrt{-\Theta} F_{\mu\nu}}\Big)$ at leading order in the derivative expansion (which implies that the action depends also on $\tilde F_{\mu\nu} F^{\mu\nu}$, but we are ignoring this here). What we want to highlight now is that the iDBI case is also protected by the same symmetry, as it is quite obvious since the structure of the Lagrangian is the same. In the geometrical realization, the $\Theta>0$ case represents  a brane with negative tension yet with a healthy gauge field embedded in it, which corresponds just to the continuation of \eqref{powermodel} with $p=1/2$ to $\Theta>0$. The negative tension might raise concerns about the potential problems finding a UV completion of this action. Some of these concerns have been rather sharply articulated previously, {\em e.g.}, in \cite{Adams:2006sv}: EFTs with higher order operators with the `wrong' sign (such as what happens in the iDBI case) are argued to present generically super-luminal modes and therefore to lack a Lorentz invariant UV completion. Let us note only here that it is quite obvious that the UV completion of the low-energy effective scale-invariant field theory that captures strongly coupled materials is inevitably non-relativistic (at least in its first step, which must be phrased in terms of non-relativistic degrees of freedom -- atoms). It is at present unclear to us whether or not the iDBI models have superluminal modes (see the Appendix A for some discussion). But even if there were, it is not clear either whether this is directly a problem. {In the effective holographic models the light-cone structure of the gravity dual corresponds to an emergent light cone, which in realistic cases has a subluminal limiting velocity (such as what similarly happens in graphene)}. Therefore, we don't find that this is an immediate logical obstacle that forces one to disregard the models of iDBI type, even though it is certainly an issue that deserves further study.  

In any case, whether or not the choice of NED Lagrangian is protected by a symmetry (and whether or not it admits a Lorentz invariant UV completion) is a secondary issue for our analysis, so we will just continue with a generic $K$ and particularize for the two models \eqref{models} and show the results. 
In the sequel, we analyze the implications of introducing such non linearities in the context of the transport properties of the dual CFTs.

\section{Electric response}\label{sec3}
\subsection{Electric conductivity}
The DC conductivity for this model reads:
\begin{equation}
\boxed{\sigma_{DC}=K'(\PPi^2(u_h))+\frac{q^2\,\rho^2 \,u_h^2}{M^2(u_h)}}
\label{DCformula}
\end{equation}
where $M^2(u_h)=m^2\,\alpha^2\,V'(\alpha^2\,u_h^2)$. The interested reader can find the derivation of this result in Appendix \ref{app2}. For future reference, we will define the first term in \eqref{DCformula} as the `probe' conductivity (see Eq.~\eqref{ProbeDC}),
$$
\sigma_P=K'(\PPi^2(u_h))~.
$$
For the benchmark models \eqref{models} along with the choice of a linear potential 
$$
V(z)=z/2m^2
$$ 
for the $\phi$ scalars (so that the $\alpha$ parameter encodes the disorder strength), the DC conductivity reads:
\beq   \label{transport}
\sigma_{DC}\,=\,\sqrt{1-\rho ^2\, u_h^4\,\Theta}+\frac{2 \,q^2\,\rho ^2 \,u_h^2}{\alpha^2}\,.
\eeq
From this and \eqref{T},  one can straightforwardly obtain how $\sigma_{DC}$ depends on $T$, $\rho$, etc. In this section we focus on the iDBI models \eqref{models} with full backreaction, since this gives rise to an insulating behaviour.

\begin{figure}
\centering
\includegraphics[width=4.5cm]{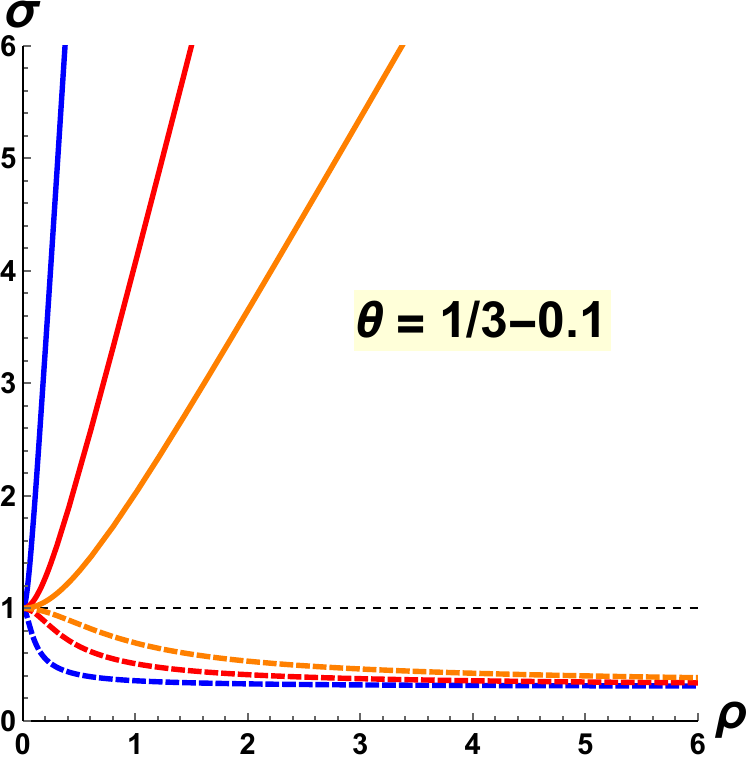}%
\;
\includegraphics[width=4.5cm]{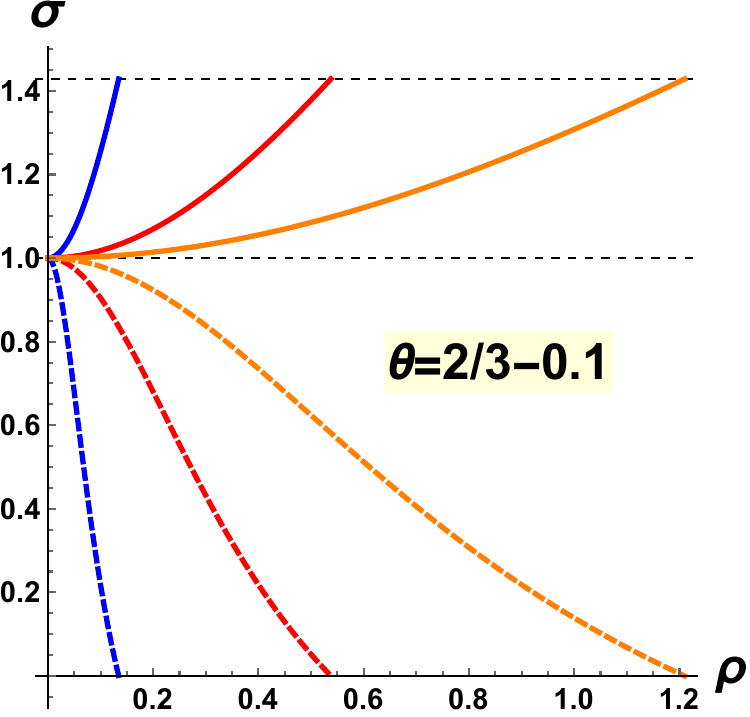}%
\;
\includegraphics[width=4.5cm]{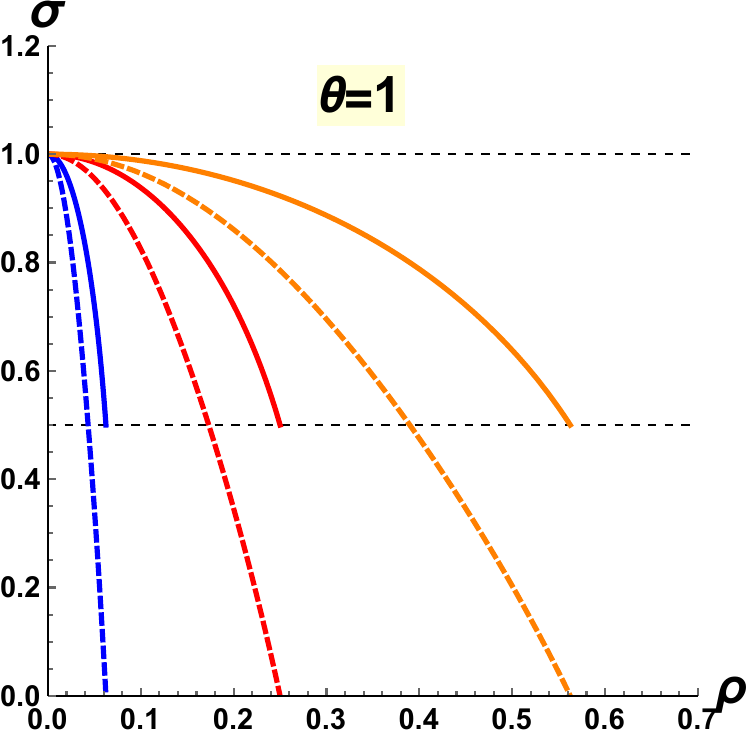}%
\caption{Probe \eqref{ProbeDC} (dashed) and full \eqref{transport} (solid) DC conductivity at zero temperature in function of the charge density $\rho$ for the various regimes of $\theta\equiv\Theta/q^2$. One can see that for $\theta>2/3$ (bottom-right) the DC conductivity decreases increasing the charge density and gets a minimum value which goes to zero $\propto 1/\theta$.}
\label{probesigmatot}
\end{figure}

\begin{figure}
\centering
\includegraphics[width=5cm]{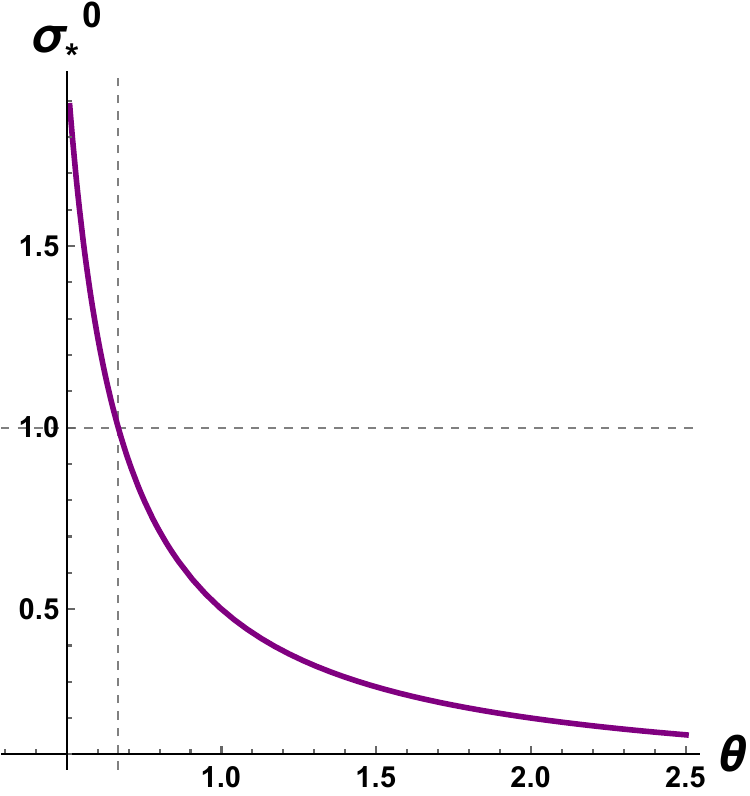}%
\caption{$\sigma^0_\star$, the DC conductivity at $T=0$ and $\rho_{\star}$, as given by \eqref{sigmastar} in function of $\Theta$. For simplicity we fix $q=1$. Note this quantity exists only for $\Theta>q^2/3$.}
\label{probesigmatot2}
\end{figure}

It is quite clear upon series expansion around $\rho=0$ that there is a very dramatic change of behaviour for \eqref{transport} at $T=0$ above a certain threshold for the self-interaction parameter, specifically $\Theta>(2/3) q^2$ \footnote{This defines the second threshold mentioned above as
$\Theta_2 = 2 q^2 /3$.
}. Above this threshold, the DC conductivity decreases with increasing $\rho$ (at constant $T$), a feature that is very far from the Drude metal. Instead, this is quite reminiscent of the traffic jam picture of a Mott insulator.
Note that this threshold is larger than the threshold $\Theta>q^2/3$ that imposes an upper bound on the  the charge density $\rho<\rho_\star$. 
The various different bahaviours of the electric conductivity (at zero temperature) that one can have depending $\Theta$ are summarized in Table \ref{tab:table1} and in Figs.~\ref{probesigmatot} and \ref{probesigmatot2}.

More specifically, for non-zero disorder strength $\alpha$ one can distinguish 4 cases:
\begin{itemize}
\item $\Theta <  q^2 /3$ : there is no problem to satisfy $u_h<u_\star$ at any $T$ and $\rho$. The probe conductivity at $T=0$, $\sigma_{P}^0$, \eqref{ProbeDC} is positive and less than 1 ( the Maxwell value in unit of $q^2$); the complete one at $T=0$, $\sigma_{DC}^0$ is positive too and getting larger and larger than 1 increasing the charge density $\rho$.
\item $q^2/3<\Theta <  2\,q^2 /3$ : for $\theta>q^2/3$ there is always a maximum density, $\rho_{\star}$, such that $u_{\star}=u_0$ and $\sigma_{P}^0=\chi^0=0$ (probe DC conductivity and charge susceptibility at zero temperature), with $\rho_\star$ given in \eqref{rhostar}. The full DC conductivity (given by \eqref{transport}) at this particular value of the charge density and at zero temperature reads:
\beq\label{sigmastar}
\sigma_{DC\star}^0=\frac{q^2}{3\,\theta-q^2}
\eeq
and it is surprisingly independent of the disorder strength $\alpha$.
In this regime, $\sigma^0_{DC}$ satisfies $\partial \sigma^0_{DC}/\partial \rho >0$ (no traffic jam). At $\rho_{\star}$,  this is larger than 1, but bounded.
\item $\Theta =  2\,q^2 /3$ : At $\rho_{\star}$ we have $\sigma_P^0=0$ while $\sigma^0=1$ (in units of $q^2$) for every $\rho_\star$.
\item $\Theta \geq  2\,q^2 /3$ : At $\rho_{\star}$ we have again $\sigma_P^0=0$. Now, though,   there {\em is} `traffic jam', $\partial \sigma_{DC}^0/\partial \rho <0$. The value of $\sigma_{DC}^0$  at the maximum density $\rho_{\star}$ goes like  $q^2/(3\,\Theta)$ for large $\Theta$ and can be therefore arbitrarily low.
\end{itemize}

\begin{table}[h!]
  \centering
  \label{tab:table1}
  \begin{tabular}{ccccc}
    \toprule
     $\theta\equiv\Theta/q^2$ & bound on $\rho$ & $\sigma_{P\star}^0$ & $\sigma^0_{DC\star}$ & $\partial\sigma_{DC}^0/\partial \rho$\\
    \midrule
    $\theta< 1/3$ & no bound  & $1>\sigma_{P}^0|_{\rho\to\infty}>0$ & $\sigma_{DC}^0|_{\rho\to\infty}\to\infty$ & $>0$\\[2mm]
    $1/3<\theta<2/3$ & $\rho\leq \rho^{\star}$ & $\sigma_{P\star}^0=0$ & $\sigma_{DC\star}^0=\frac{1}{3\,\theta-1}>1$  & $>0$\\[2mm]
    $\theta=2/3$ & $\rho\leq \rho^{\star}$ & $\sigma_{P\star}^0=0$ & $\sigma_{DC\star}^0=1$ & $0$\\[2mm]
    $\theta>2/3$ & $\rho\leq \rho^{\star}$ & $\sigma_{P\star}^0=0$ & $\sigma_{DC\star}^0=\frac{1}{3\,\theta-1}<1$ & $<0$\\
    \bottomrule
  \end{tabular}
  \vspace{0.3cm}
      \caption{DC electrical conductivity $\sigma_{DC}$ at zero temperature for the iDBI model ($\theta>0$). The full conductivity at $T=0$ is bounded between $1$ and $\sigma_{DC\star}^0$. The `probe' conductivity at $T=0$ (defined as $\sigma_P^0=K'|_{u^0}$), is bounded between 1 and $\sigma_{P\star}$, which is the limiting \color{black} value at the maximal density $\rho=\rho_\star$. 
  }
\end{table}

~\\~\\

We now turn to the study of the `optical' AC conductivity for the iDBI model in different setups and we will come back to the $T=0$ features in the next section \ref{MITsec}. For the rest of the paper we fix $q=1$ unless explicitly stated.

\begin{figure}[htbp]
\centering
\includegraphics[width=42mm]{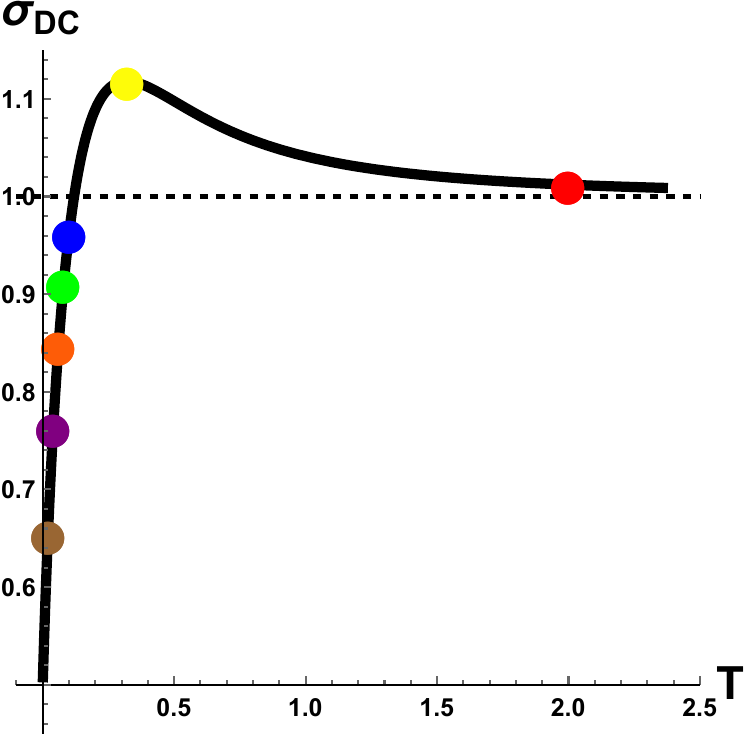}%
\qquad
\includegraphics[width=45mm]{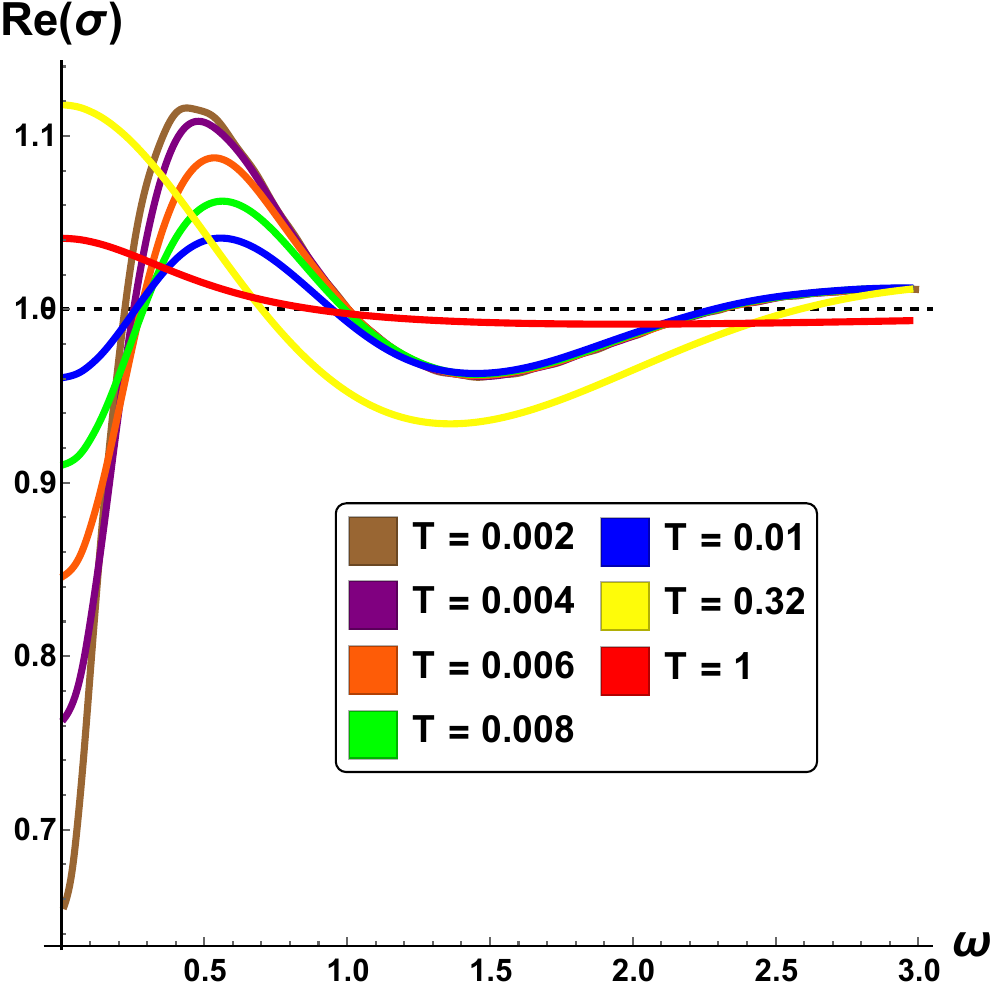}%
\qquad
\includegraphics[width=45mm]{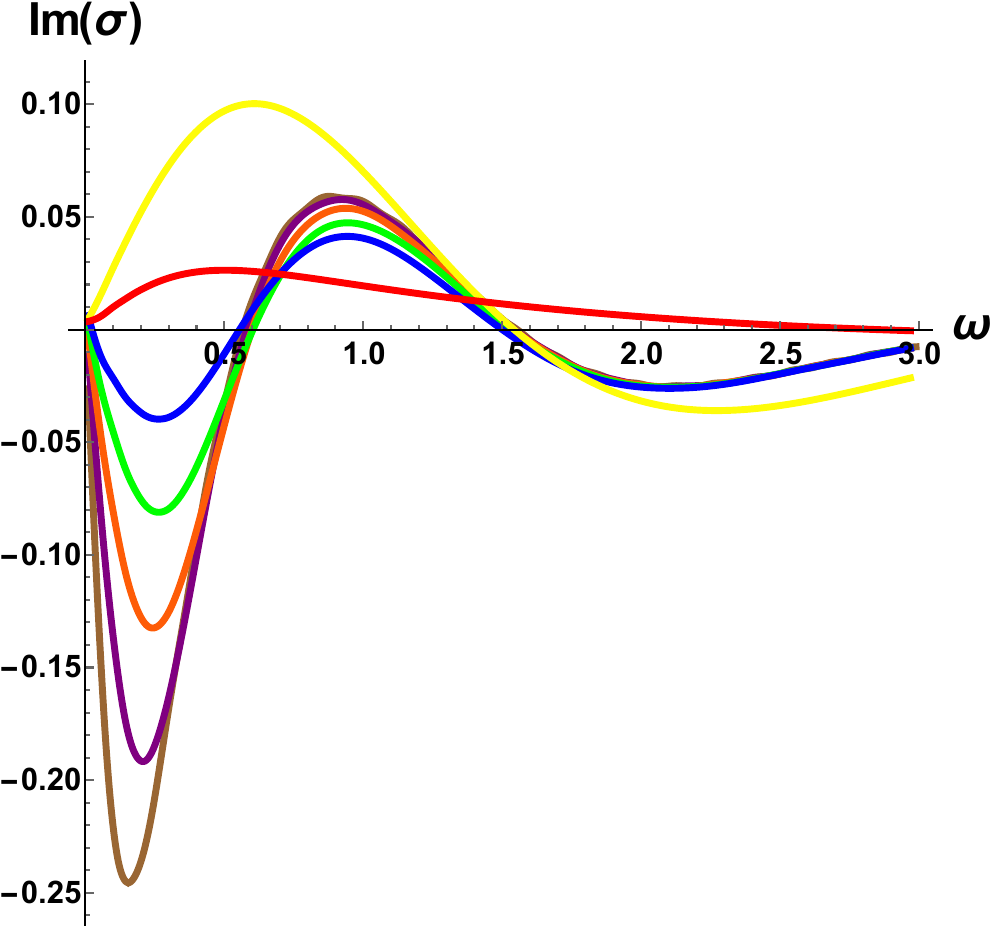}
\caption{Transport properties for the backreacted iDBI \eqref{models} with $\Theta=1$ and $V(z)=z$ and: $m^2=\rho=1.8,\,\alpha=\sqrt{2}$.}
\label{ACback1}
\end{figure}

The first case we present is the linear metric potential case $V(z)= z$ (the model of \cite{Andrade:2013gsa}), which in the absence of non linear electronic interactions has metallic behaviour. On top of this background we switch on the iDBI non-linearities in the electron sector which results in  a suppression of the DC conductivity. The results are shown in Fig.~\ref{ACback1} for some representative parameters.

The iDBI case with $V(z)=z$ already shows quite remarkable features:
\begin{itemize}
\item It suppresses the DC conductivity, leading to  insulating behaviour at low T.
\item The competition between the linear potential for the bulk phonons $\phi$ and the non linear iDBI for the electrons produces already a \textit{metal-insulator crossover} upon dialing the temperature of the geometry.
\item A pinned response in the optical conductivity appears -- it is entirely due to the electron sector. 
For small $T$, we notice the appearance of a rather broad but certainly visible resonance at nonzero frequencies. This resonance becomes narrower for bigger $\Theta$ and/or $\rho$ closer to $\rho_\star$. Given the interpretation that we are proposing here as traffic-jam-like behaviour, it is natural to interpret this resonance as an {\em accordion wave} effect.
\color{black} 
\item Even at temperatures where the Drude peak is present it seems that also additional features appear at higher-frequencies.
\end{itemize}

\begin{figure}[htbp]
\centering
\includegraphics[width=40mm]{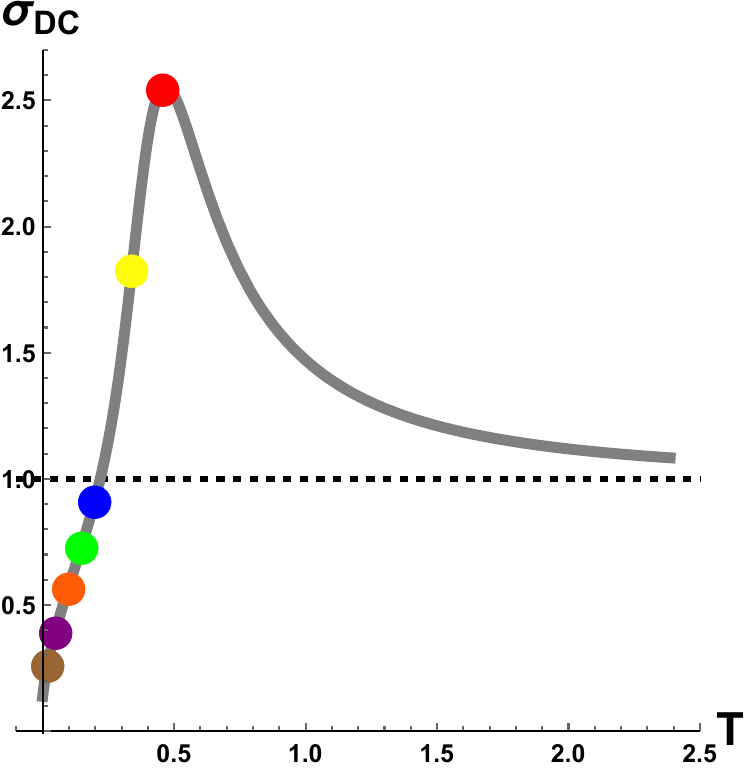}%
\qquad
\includegraphics[width=45mm]{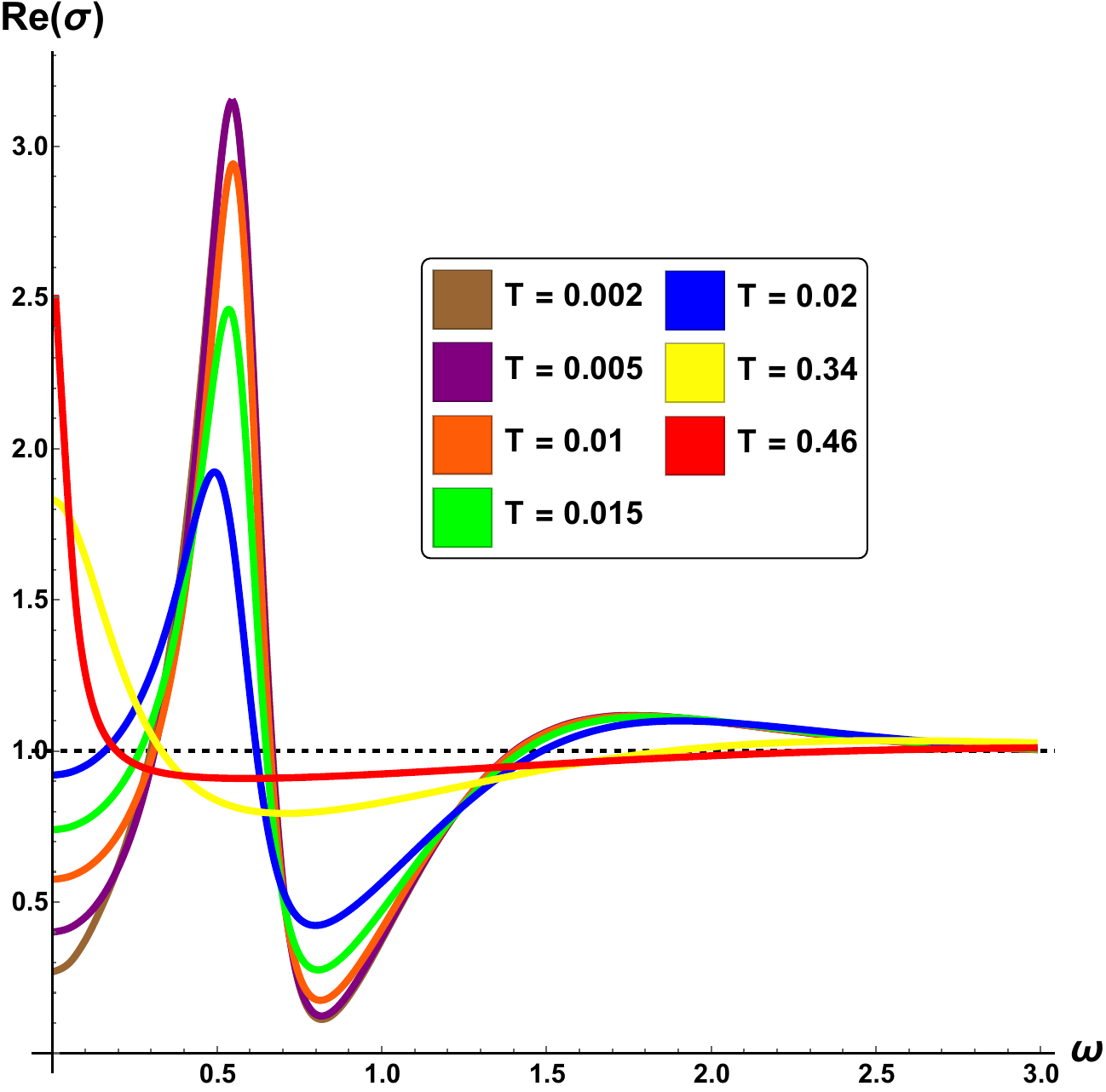}%
\qquad
\includegraphics[width=45mm]{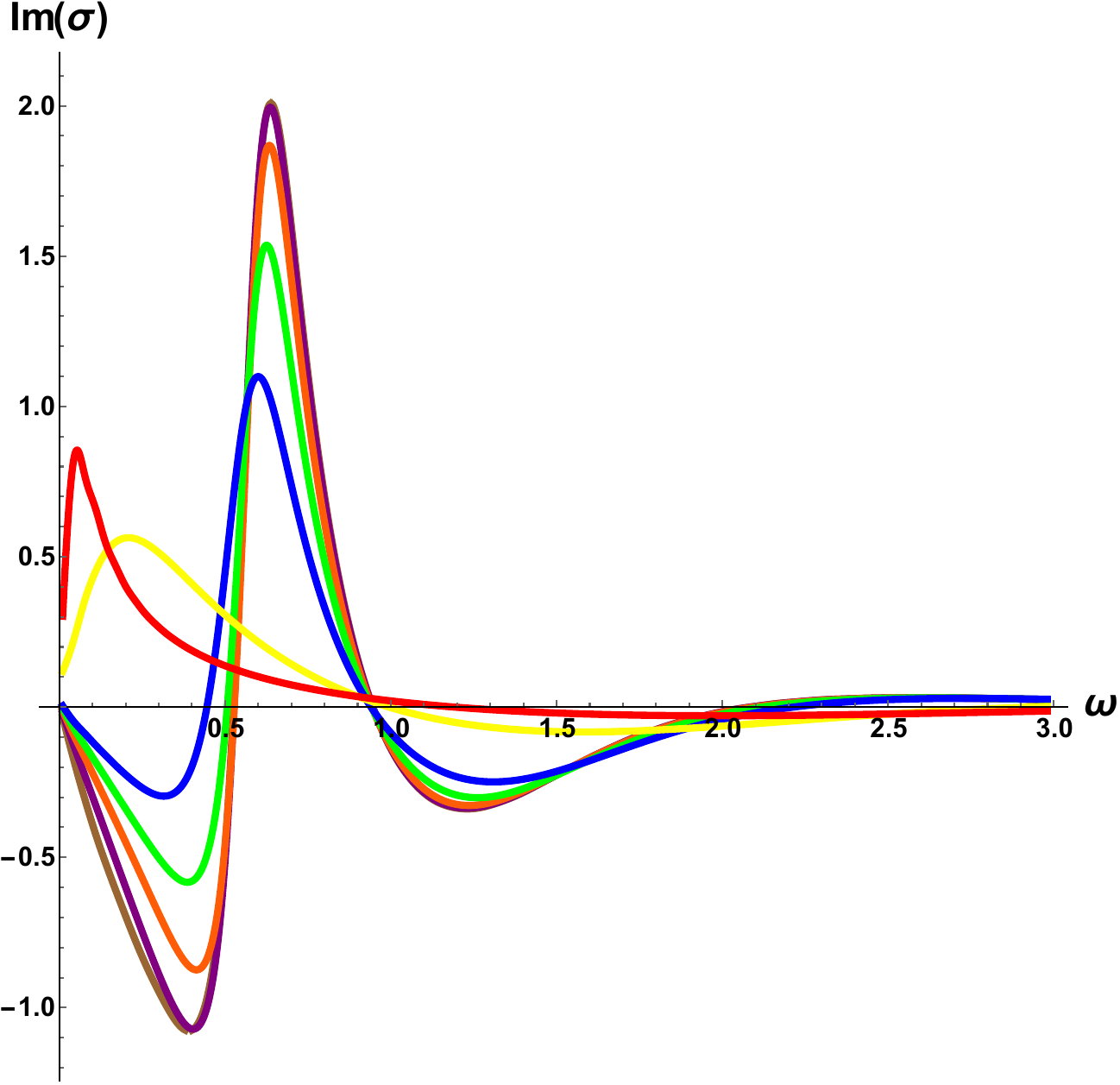}
\caption{Transport properties for the backreacted iDBI with non linear potential $V(z)=z+z^5$ and $\Theta=1$.
Parameters are: $m^2=0.06,\,\rho=1\,,\alpha=\sqrt{2}$.}
\label{ACback2}
\end{figure}

One can of course complicate the situation considering also non linear potential in the bulk phonons sector. One interesting case is for example to take the potential of \cite{Baggioli:2014roa}.
The results for the iDBI model with non linear potential $V(z)=z+z^5$  are shown in figure \ref{ACback2}, and can be summarized as:
\begin{itemize}
\item The  DC conductivity at low temperature is considerably reduced, thus realizing a rather good insulator  ($\sigma_{DC}(T=0)\ll1$) and keeping the metal-insulator crossover already present in the model at finite $T$.
\item The mid-infrared peak observed in \cite{Baggioli:2014roa} is still present in the optical conductivity and it seems to be enhanced by the iDBI non linear extension, meaning that electron-electron interactions can  contribute to sharpen that resonance. 

Since in this case the resonance is also present at $\Theta=0$, this resonance represents a polaron \cite{Baggioli:2014roa} which, at $\Theta\neq0$, also includes charge-sector nonlinearities.
\color{black}
\item The iDBI non linear interactions provide a second mild peak for larger frequencies which was not present in  \cite{Baggioli:2014roa}. It would be interesting to understand further the nature of that peak in relation to the pinned response of real Mott Insulators. 
\end{itemize}

\subsection{Metal-Insulator transitions}\label{MITsec}

The next interesting question is: can we find out a model where the non-linearities in the electronic sector (e-e interactions) provide a metal-insulator transition?  Is there a tunable parameter we can dial to drive a MIT?
In other words, can one find a \textit{Quantum Phase transition} where the two phases (metallic and insulating) are connected dialing an external parameter while keeping the temperature to be zero?

\begin{figure}[htbp]
\centering
\includegraphics[width=95mm]{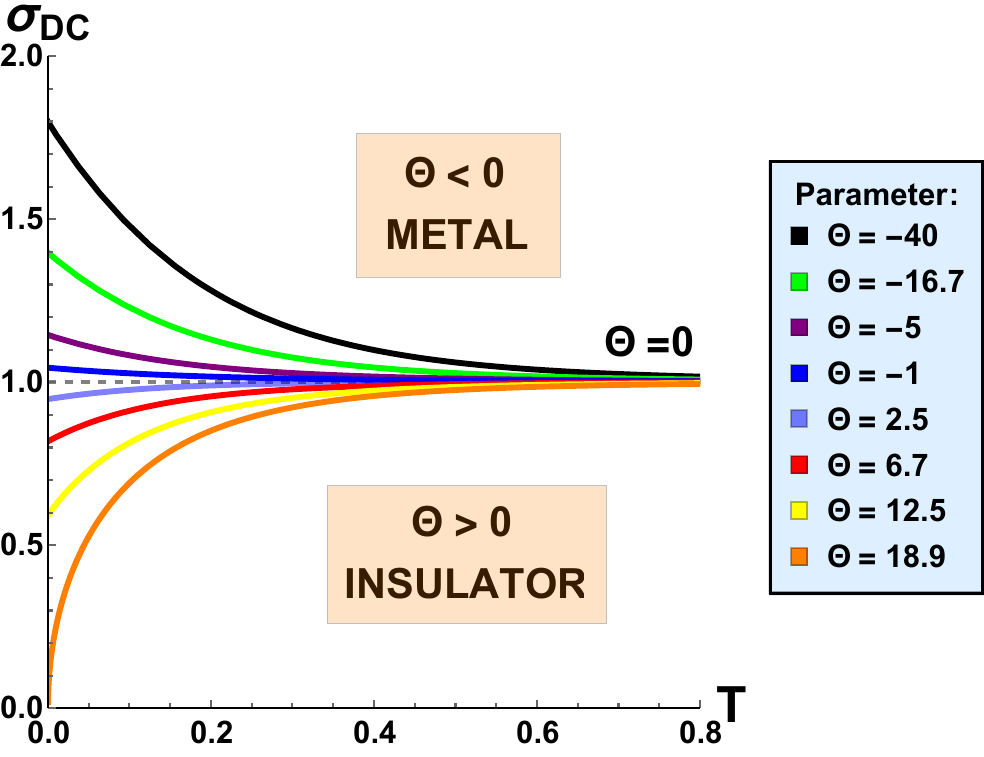}
\caption{An example of \textit{Mott transition} in our model. The \textit{Metal-Insulator Transition} is obtained dialing the $\Theta$ parameter of the model \eqref{powermodel} with $p=1/2$. The potential for the $\phi$ scalars is taken to be linear $V(z)=z/2m^2$. $\rho=0.4$ and $\alpha=3.2$ .}
\label{mitgamma}
\end{figure}

It is then necessary and interesting to find out a tunable parameter whose dialing can induce such a mechanism. Taking into account that the  iDBI corresponds to an insulator while the DBI to a metal, it quite obvious that that the parameter  connecting those models in a continuous way is none other than the $\Theta$ parameter that controls the strength (and sign) of the e-e interactions in the benchmark models \eqref{models}.
Fixing the charge density and the disorder strentgh (i.e. graviton mass) and dialing the $\Theta$ parameter we can indeed provide in the full backreacted model a nice metal-insulator transition which is clearly driven just by the electronic sector's non linearities (that is, e-e interactions) and it represents a holographic simple toy  example of the so called \textit{Mott transition}. An example of this result (making use of formulas \eqref{transport}) is shown in fig.\ref{mitgamma}.

As expected, we can interpolate between the metallic and the insulating behaviour dialing $\Theta$, that is,  by changing the non-linear terms in the field strength $F_{\mu\nu}$. We can indeed focus on the $T\approx 0$ value of the DC and see how this goes from a finite (i.e. $\sigma>1$) value to a small and eventually $\approx 0$ one dialing $\Theta$. $\Theta=0$ corresponds to the Maxwell case; once we start increasing the value of $\Theta$ we depart from the Maxwell case and depending if we go towards negative or positive values we run into a metallic or an insulating phase.

\begin{figure}[htbp]
\centering
\includegraphics[width=58mm]{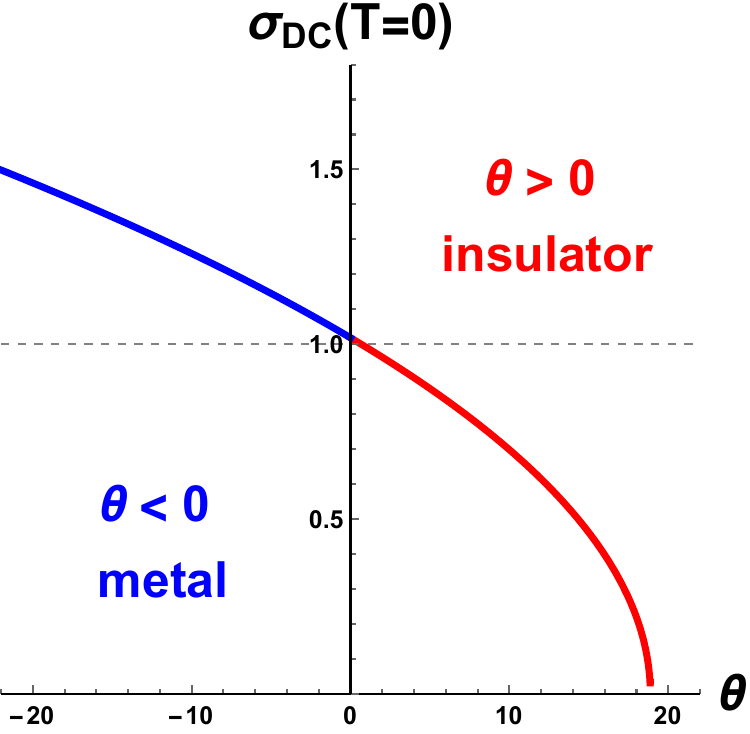}
\caption{Electric DC conductivity at zero temperature obtained dialing the $\Theta$ parameter of the model \eqref{powermodel} with $p=1/2$. The potential for the $\phi$ scalars is taken to be linear $V(z)=z/2m^2$. The left parameters are fixed to be $\rho=0.4$ and $\alpha=3.2$. The minimum DC value reads $\approx 0.017$. For the full T behaviour see fig.\ref{mitgamma}.}
\label{DCT0plot}
\end{figure}

From the technical point of view the extremality condition (once fixed the amount of charged density $\rho$ and the disorder strength, i.e. the graviton mass, in an safe way following fig.\ref{probesigmatot1}) fixes a maximum value for $\Theta=\Theta_0$. The DC conductivity at that value can be computed analytically from the expression \eqref{transport}. Depending on the value of $\rho$ and $\alpha$ the DC value at $T=0$ (namely at $\Theta=\Theta_0$) can become arbitrarily small (but exactly zero just at $\rho=\rho^{\star}$ and $\Theta=\infty$, see table \ref{tab:table1}).
 \\This certainly requires a certain tuning of parameters\footnote{The notion of tuning in condensed matter is slightly different than in particle physics. Most of the High $T_c$ superconductors are synthetic compounds that present significant amounts of design, and therefore can be considered as tuned systems.}, as we did in the example of Fig.\ref{mitgamma}.
To make things clearer we also plot the value of the DC conductivity at zero temperature  in Fig.\ref{DCT0plot}, with the same parameters of Fig.\ref{mitgamma}.

Let us emphasize is that at this level, there is no direct link of the $\Theta$ parameter into external control parameters that can be used in real condensed matter, and this question is beyond the scope of this article. One clear statement, though,  is that $\Theta$ clearly has to do with  the amount (and sign) of non-linearities in the Maxwell sector. Dialing $\Theta$, then, must correspond to increasing/decreasing the strength of electron self-interactions and consequently their mobility. We expect the iDBI case to mimic some sort of strong coulomb interaction between electrons which leads to pinning behaviour \`a la Mott. Definitely, it would be illuminating to understand better its role and search for a better parameter to dial to compare with real situations.

Let us also mention that these results also imply that same model also exhibits MITs (at fixed $\Theta$) by varying various other external parameters such  as $\mu$, $\rho$ or $T$, at least in some region of the parameter space (and with the conductivity decreasing in the insulating phase to a finite but small value). 
\color{black}

\subsection{Non-linear susceptibility}

\begin{figure}
\centering
\includegraphics[width=13.5cm]{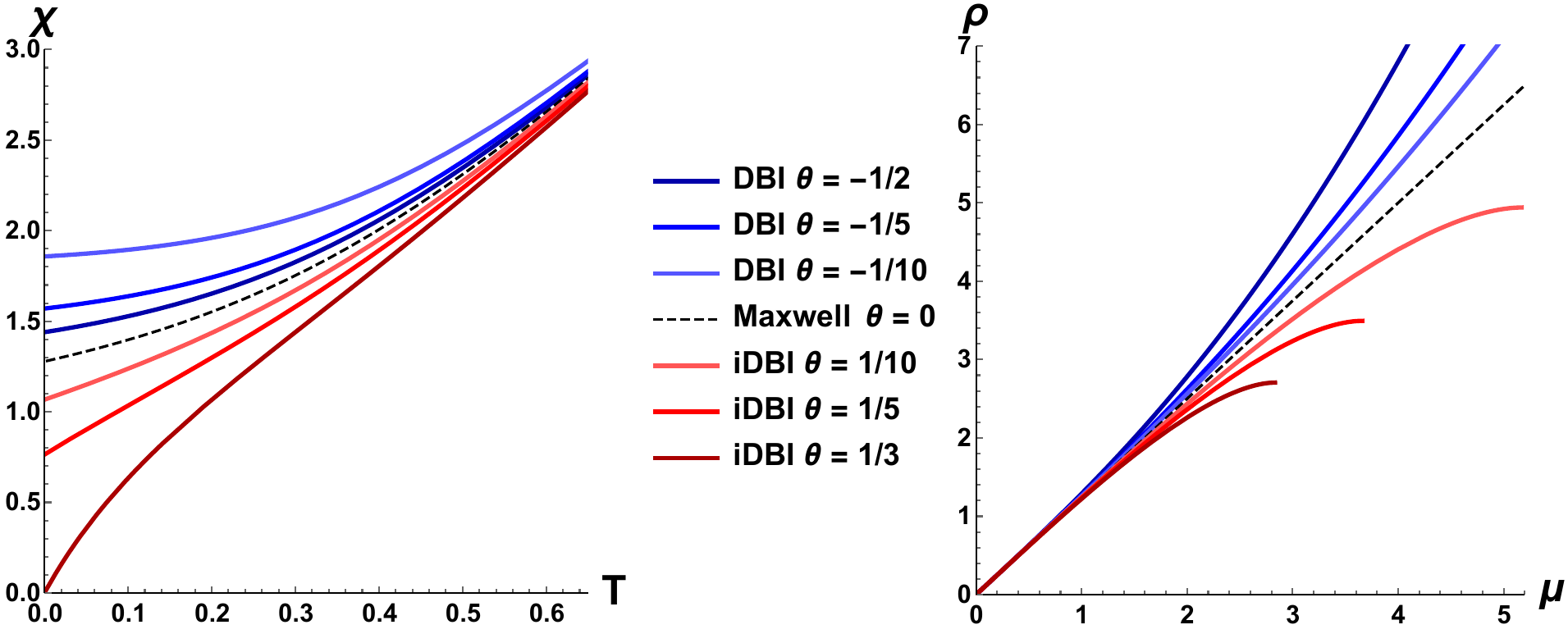}
\caption{Non linear susceptibility for the models \eqref{models} with $\alpha=0$. \textbf{Left:} Static charge susceptibility $\chi$ in function of temperature T at fixed charge density $\rho=4$; \textbf{Right:} $\rho-\mu$ curve at fixed $u_h=0.8$. Clearly, the iDBI case presents nonlinear screening (a reduction of $\rho$ compared to the linear extrapolation). Conversely the DBI case presents nonlinear anti-screening. Note that both at big temperature T and small chemical potential $\mu$ the non linear models reduce to the linear Maxwell case. Note also that for $\Theta=1/3$ the susceptibility $\chi$, like the probe conductivity $\sigma$, goes to 0 at null temperature (see also fig.\ref{probesigmatot}).}
\label{CSuh}
\end{figure}

From the non-linear nature of the charge sector of these models, it it clear that a very generic property of these models is that they should display a strong nonlinear response. Let us initiate here the analysis of the nonlinear response by studying the simplest in principle observable, namely, the \textit{static charge susceptibility}. See also  \cite{Nogueira:2011sx} for similar  analyses  in the holographic context.\\
The solution for the gauge field in the general model reads:
\begin{equation}
A_t(u)\,=\,\int_{u}^{u_h}\frac{\PPi(t)}{t^2}\,dt
\label{Amu}
\end{equation}
and already implements the regularity condition at the horizon.
From \eqref{Amu} we can define the chemical potential and the charge density for instance as
\begin{align}
& \mu\,=\,A_t(0)-A_t(u_h)\,,\,\,\,\,\,\,\,\,\rho\,=\,-A_t'(0)\,,
\end{align}
which is as much as requiring the gauge field to asymptote close to the boundary to $A_t=\mu-\rho\, u$.
For the Maxwell theory the solution reads $\PPi(u)=\rho\, u^2$ such that we recover the usual expression $\mu=\rho \,u_h$. Consequently, the \textit{static charge susceptibility} (at constant entropy density $s\,=\,2\pi/u_h^2$) 
$$
\chi_E\,=\frac{\partial\rho}{\partial\mu}
$$ 
is just equal to $u_h^{-1}$.
Instead, in the non-linear models \eqref{models} the relation between chemical potential and charge density instead is easily seen to be given by
\begin{equation}
\mu\,=\,\rho\,u_h\;{}_{2}F_1\left(\frac{1}{4},\frac{1}{2};\frac{5}{4};u_h^4\,\rho^2\,\Theta\right)~,
\end{equation}
where ${}_{2}F_1\left(a,b;c;z\right)$ is the hypergeometric function. 
Again we see that at large temperature, which corresponds to small $u_h$, these theories are just a small deformation around the linear Maxwell theory:
\begin{equation}
\mu\,=\,\rho\,u_h\,+\,\Theta\,\frac{\rho ^3\,u_h^5}{10}+O\left(u_h^6\right)+\dots
\end{equation}
An example of susceptibility in function of temperature (at constant charge density) and of the $\rho-\mu$ curve at constant entropy density is shown in figure \ref{CSuh} for different $\Theta$ and zero momentum dissipation.
\begin{figure}
\centering
\includegraphics[width=4.2cm]{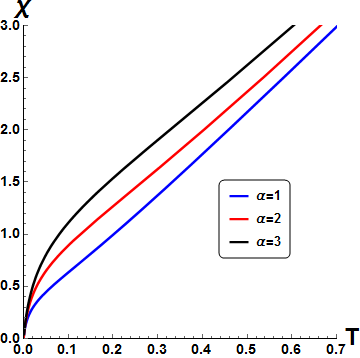}%
\qquad
\includegraphics[width=5cm]{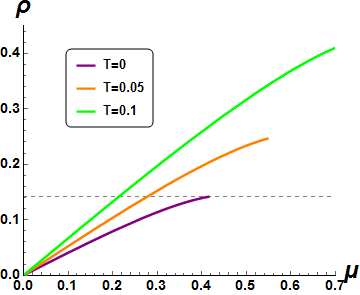}%
\qquad
\includegraphics[width=4.2cm]{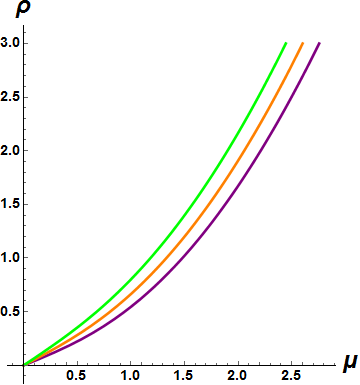}%
\qquad
\caption{More on the susceptibility $\chi$ for the iDBI model (q is fixed to 1 everywhere). \textbf{Left:} Susceptibility in function of temperature for different momentum strengths $\alpha$ at $\rho=\rho^{\star}$  and $\Theta=2$. Note as $\chi(0)=0$ which is indeed a generic fact at $\rho=\rho^{\star}$ and it correlates with a vanishing probe DC conductivity. \textbf{Center:} $\rho(\mu)$ for $\Theta=2$, $\alpha=1$ and different temperatures. Note since $\Theta>q^2/3$ there is a maximum $\rho$ and $\mu$. \textbf{Right:} $\rho(\mu)$ for $\Theta=0.2$, $\alpha=1$ and different temperatures. Note since $\Theta<q^2/3$ there is no maximum $\rho$ and $\mu$.}
\label{moresus}
\end{figure}
One can see the two different effects induced by the non linear \eqref{models} models are sensitive just in the range of large $u_h$ (small $T$), as expected.  While the DBI model enhances the response of the charge density to a chemical potential, the iDBI decreases it. 
More about this quantity is shown in fig.\ref{moresus} where the plots are generalized for finite momentum dissipation and at constant temperature T.

The common feature is that at $\rho=\rho^{\star}$ (and at $\Theta=q^2/3$ for zero momentum dissipation) the charge susceptibility $\chi$ is zero at zero temperature and this correlates with the vanishing of the probe conductivity.
As already shown in fig.\ref{probesigmatot1} for $\Theta>q^2/3$ there is a maximum amount of charge density $\rho^{\star}$ (and of relative chemical potential $\mu^{\star}$) which the system can afford, while for $\Theta<q^2/3$ there is no such a bound as shown in fig.\ref{moresus}.




\section{Superconducting instability}\label{SCsection}
Strongly correlated electronic systems provide for a richer and  more complicated phase diagram  than the one we discussed so far. Upon varying parameters like the temperature or the doping (mobile charge density, {\em i.e.}, $\rho$), new phases arise -- most notably a \textit{superconducting} (SC) phase. 
The SC phase shows up in a dome-shape region in the $T-\rho$ phase diagram as sketched in Fig.~\ref{SCsketch} and transitions between the SC phase and the \textit{normal} phase (metallic or insulating) appear in the form of Quantum Phase transitions (at $T=0$) still affecting a wider region of the actual phase diagram which takes the name of \textit{quantum critical region} (QC).

The SC phase transition in these strongly correlated systems can be modeled  using the tool of the AdS-CFT correspondence \cite{Hartnoll:2008vx,Hartnoll:2008kx} (see also \cite{Baggioli:2015zoa,Kiritsis:2015hoa,Baggioli:2015dwa,Chen:2016cym} for some recent progress towards reproducing this phase diagram
in the context of  effective holographic theories).
Previous studies of the Insulator-Superconductor transition in holography include \cite{Nishioka:2009zj,Jing:2012dj}.
\color{black}
The way to account for the SC transition in effective holographic models is by realizing that the charged fermion bilinear $\OO_{SC}$ that can condense (and thereby break the electromagnetic $U(1)$ symmetry) is a charged scalar operator. Therefore in the gravity dual one must include that degree of freedom -- a charged scalar field $\psi$ dual to the fermion bilinear.
At the lowest order in derivatives and in powers of  $\psi$ (which is a valid approximation at low energies and near the SC transition) the effective action in the gravity dual is
\begin{equation}
\mathcal{S}_{SC}\,=\,-\,\int\,d^4x\, \sqrt{-g}\,\left(\,|D\,\psi|^2\,+\,M^2\,|\psi|^2 +\dots\,\right)
\end{equation}
with $D_\mu \psi=(\partial_\mu-i\,g\,A_\mu)\psi$  and $g$ the charge  of $\psi$.

As usual, whenever the v.e.v. of the condensate (in the ground state) is non vanishing, $\braket{\OO_{SC}}\neq 0$, the system is in the SC phase\footnote{In the gravity dual, and in the normal quantization, $\braket{\OO_{SC}}$ is encoded in the normalizable mode of the scalar field $\psi$ near the UV boundary, see \cite{Hartnoll:2008vx} for further details.}.
In the gravity dual, upon decreasing the temperature $T$ or increasing the charge density $\rho$ one expects the system to spontaneously develop a non trivial profile for the scalar field $\psi$ with $\braket{\OO_{SC}}\neq 0$ and eventually the  normal phase develops a SC instability.

In order to construct the full SC solution and phase diagram one needs to resort to numerical methods. Still, one can use a simpler criterium to identify whether the SC instability appears or not,  relying on properties of the presumed normal phase at $T=0$. At `extremality' the near-horizon geometry takes the usual form $AdS_2\,\otimes\,\mathcal{R}_2$ and a robust criterium to check the instability of the background is given by the so called BF bound violation:
\begin{equation}
M_{eff}^2\,L_2^2\,<\,-\frac{1}{4}
\end{equation} 
where $L_2$ is the size of the $AdS_2$ geometry at the extremal horizon $u=u_0$ given by:
\begin{equation}
L_2^2\,=\,\frac{2\,L^2}{f''(u_0)\,u_0^2}
\end{equation}
and $M_{eff}^2$ is the effective mass of the scalar field $\psi$ at the horizon which can be written like:
\begin{equation}\label{meff}
M_{eff}^2\,=\,M^2\,+\,g^2\,A_t^2\,g^{tt}
\end{equation}
The second term is  negative and it can grow large at the horizon and the effective mass $M_{eff}$ can go below the BF bound and produce an instability of the background towards a new solution where the profile of the scalar $\psi$ is not trivial (i.e. the SC phase).

Since the second term in \eqref{meff} involves the value of the gauge field $A_t$ at the horizon the instability is {\em very sensitive} to possible non linear structure in the charge sector like the ones introduced in the previous sections. One immediate question is therefore whether these non-linearities enhance or obstruct the SC transition. We will restrict ourselves to the case $M=0$, but other choices of $M$ do not alter much the main results. The parameter that controls whether SC appears or not then is
\begin{equation*}
\zeta_{SC}\,=\,\,A_t^2\,g^{tt}\,L_2^2
\end{equation*}
computed at the extremal horizon $u=u_0$, so that 
\begin{equation*}
\zeta_{SC}\,<\,-\,\frac{1}{4}\qquad \rightarrow \qquad \textbf{SC}~.
\end{equation*}
If  $\zeta_{SC}$ is suppressed by the non linear effects the SC phase transition gets more difficult and the relative critical temperature must $T_c$ decrease (at other quantities fixed); conversely, if $\zeta_{SC}$ is increased by the non linearities than the SC instability is favored and $T_c$ must increase.

\begin{figure}
\centering
\includegraphics[width=13.5cm]{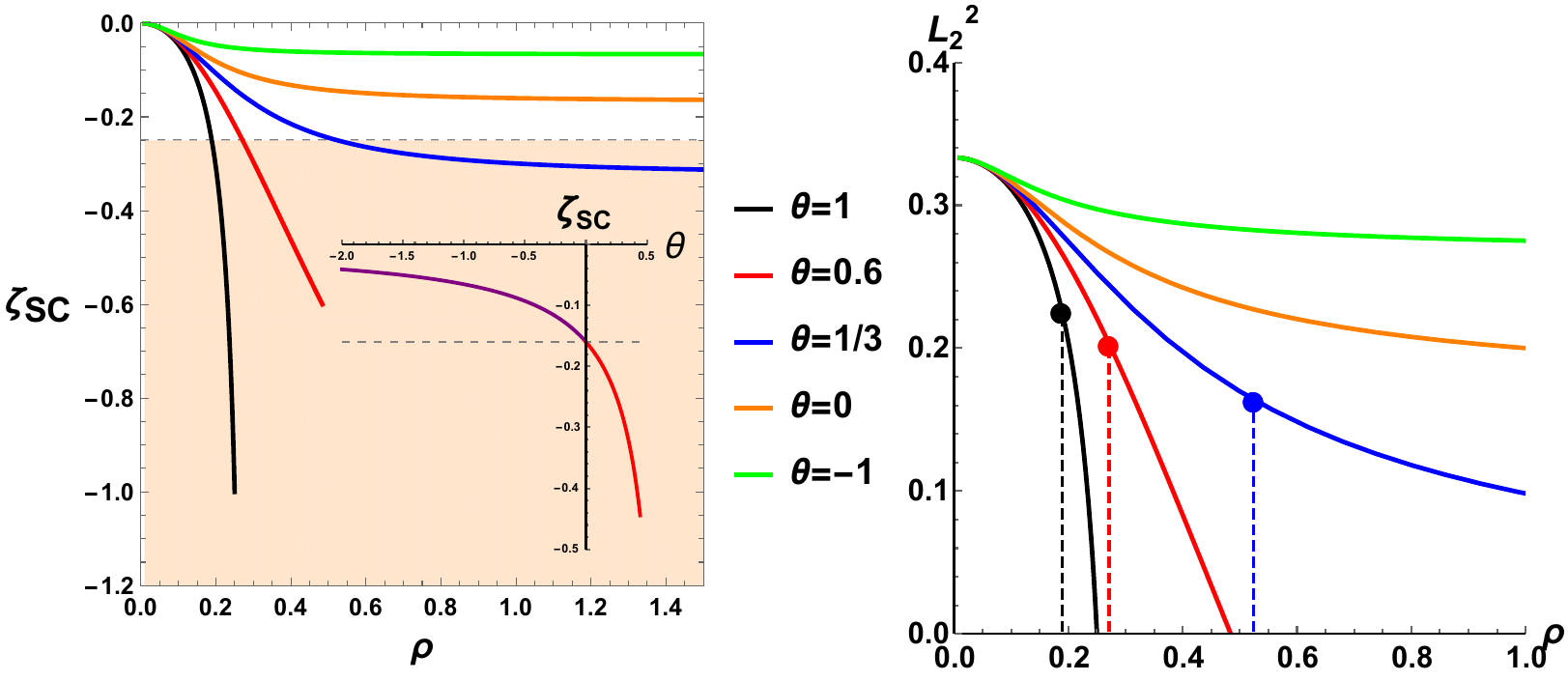}
\caption{SC instability. \textbf{Left: } the SC parameter $\zeta_{SC}$ in function of the charge density $\rho$ for $\alpha=1,q=1$ at different $\theta$. Whenever $\zeta_{SC}<-1/4$ (colored region) the SC instability appears. The inset shows $\zeta_{SC}$ at unitary charge density in function of $\theta$ and how iDBI enhances the SC instability. \textbf{Right: } the corresponding AdS$_2$ length (in units of AdS$_4$ length) for the same parameters. The dots indicate the value at which the SC instability appear for various $\theta$. Note that in the presence of a maximum charge density $\rho^\star$ the AdS$_2$ length (but not the effective dimensionless mass squared $\zeta_{SC}$) vanishes exactly at the value signaling the singularity.}
\label{SCfig}
\end{figure}

It has been shown in \cite{Gangopadhyay:2012am} that DBI extensions of the usual Holographic Superconductor model suppress the SC instability and decrease the critical temperature $T_c$. We plot the behaviour of this quantity  in Fig.~\ref{SCfig}. 

Our computation (shown \color{black} in the inset) confirms that for the DBI case $\theta<0$ the actual value of $\zeta_{SC}$ is smaller than the `Maxwell' $\Theta=0$ value, $\zeta^M_{SC}$, meaning that the SC instability is disfavored in that case. 
On the contrary, for the iDBI model $\zeta_{SC}>\zeta^M_{SC}$ indicating that the SC transition is facilitated by the non-linearities. \\

To illustrate better our results, let us introduce the critical density $\rho_{SC}$ as the value of $\rho$ where $\zeta_{SC}=-1/4$. As shown in Fig.~\ref{SCfig}, for $\Theta>0$ we find that 
\beq
\rho_{SC} < \rho_\star
\eeq
meaning that the system will develop the SC instability while the singularity is still hidden by a horizon (at $T=0$). In addition, Fig.~\ref{SCfig} also shows that the near-horizon curvature radius $\ell_2$ does not decrease too much at the SC transition (at $\rho_\star$, instead, $\ell_2$ must vanish because that is when the singularity becomes naked). \color{black} Therefore, the gravity dual does not become especially strongly coupled at $\rho_{SC}$ and so we can trust that this transition is going to happen.

One also notices form Fig.~\ref{SCfig} that $\rho_{SC}$ is quite close to $\rho_\star$ at least for large enough $\Theta$, which is when the insulator is most insulating.
From \eqref{rhostar}, we know that  $\rho_\star$ is a growing function of the disorder strength $\alpha$, and this implies that in the iDBI models the insulator-SC {\em critical doping} $\rho_{SC}$ should also be a growing function of the disorder strength (at least for sizeable $\Theta$). 
This property seems to indeed arise in recent studies of the  phase diagram for High-$T_c$ superconductors where diagram is extended on the disorder axis \cite{disdome}.

In order to make more definite statements, a proper analysis is needed of the  full phase diagram for these holographic superconductor models with iDBI non linearities, and we have to leave it for future work. In any case, the increase in $\zeta_{SC}$ seems to give a quite generic and robust reason why this class of models  should be expected to enhance perhaps dramatically the appearance of superconductivity.

\section{Decoupling limit}\label{probesection}
It is very useful to consider a decoupling or `probe' limit, where the gauge field is dynamical and non-linear but its backreaction on the metric is neglected. In this limit the dynamics simplifies significantly and one can get easily intuition and analytic control, to the extent that extracting the nonlinear response will also be quite easy. 

A sharp notion of this limit is to decouple the metric fluctuation so that the metric is `frozen' to the black brane background. Technically, this is accomplished by taking $q\to0$ and $M_P\to\infty$ keeping $q M_P$ and  $\ell$ (and $m$) finite. In this way one can focus on the non-trivial dynamics stemming from the gauge sector self-coupling. For this sector  $q$ becomes just an overall factor that does not play any role.

{This analysis can actually be done for any fixed background solution so we shall consider here that the emblackening factor $f(u)$ is an arbitrary function with a simple zero at $u_h$.
In fact, one can further distinguish various possibilities within the same decoupling limit that allows one to simplify what contributes to the background. To fix the terminology, we shall call the `probe limit I' the limit: $q\to0$ and $M_P\to\infty$ with $q M_P$,  $\ell$, $m$,  $\mu$ and $\rho$ finite completely removes the effect of the gauge sector in the metric. In this limit, the response must be insensitive to $q$. As is clear from \eqref{rhostar}, in this limit $\rho_\star ={\alpha^2  \over 6 \sqrt{\Theta} } ~.$

In addition, let us call `probe limit II', the following one: $q\to0$ and $M_P\to\infty$ with $\mu, \rho \to \infty$ keeping 
$q M_P$,  $\ell$, $m$, and  $q\,\mu$ and $q\,\rho$ finite so that we still decouple the metric and gauge perturbations but keeps a finite backraction from the charge density in the background (in $f(u)$). This limit has the advantage that one can then remove the translation-breaking sector (setting the graviton mass $m=0$)  while the DC conductivity continues to be finite. In this limit, the response must be insensitive to $m$. Indeed, Eq.~\eqref{rhostar} implies that there is no maximal value for $\rho$ but there is a bound for the self-coupling, $\Theta<(q\ell)^2/3$. From the CFT perspective, in this limit there is no momentum density operator in the dual CFT so the DC conductivity can be finite even preserving translational symmetry. 

}

\subsection{Linear conductivity}

The equation for a probe vector on top of our geometry (with unspecified $f(u)$) at zero momentum reads:
\begin{equation}
\frac{\PPi(u)}{u^2}\p_u\left(\frac{u^2\,f(u)}{\PPi(u)}a'(u)\right)+\frac{\omega^2}{f(u)}a(u)\,=\,0
\end{equation}
In this approximation the DC conductivity is just:
\begin{equation}
\sigma_{DC}^{P}=K'(\PPi^2(u_h))
\label{ProbeDC}
\end{equation}
As expected in the case of the linear theory (i.e. Maxwell), this quantity is trivially constant and it coincides with the conformal value 1.

We want to study the model \eqref{models} and its transport properties; we are particularly interested in the case iDBI cases (\ref{models}) where the dual CFT exhibits insulating behaviour\footnote{The DBI case has been extensively studied previously -- it shows purely metallic nehaviour \cite{Karch:2007pd}.}.
The DC conductivity takes the simple form
\begin{equation}
\sigma_{DC}^{P}\,=\,\sqrt{1-\rho ^2 \,u_h^4\,\Theta}~.
\end{equation}
To discuss the temperature dependence of this one needs to assume something for $f(u)$, and how it is affected by the chemical potential.
We start considering the non linearly charged BH geometry with zero momentum dissipation, $\alpha=0$, ({probe limit II}); we emphasize that in that case we necessitate $\Theta\leq (q\ell)^2/3$ to have a well defined extremal black hole. From now on, we set $q\ell=1$.
One important feature that determines the nature of the dual CFT is the DC conductivity at zero temperature which within this easy approximation can be computed directly and reads:
\begin{equation}
\sigma_{DC}^0\,=\,1 - \,3 \,\Theta
\end{equation}
It is nice to see that there is another clear indication that the model for $\Theta>1/3$ is not safe in the probe limit which is encoded in the fact that the DC conductivity becomes negative.
\begin{figure}
\centering
\includegraphics[width=6cm]{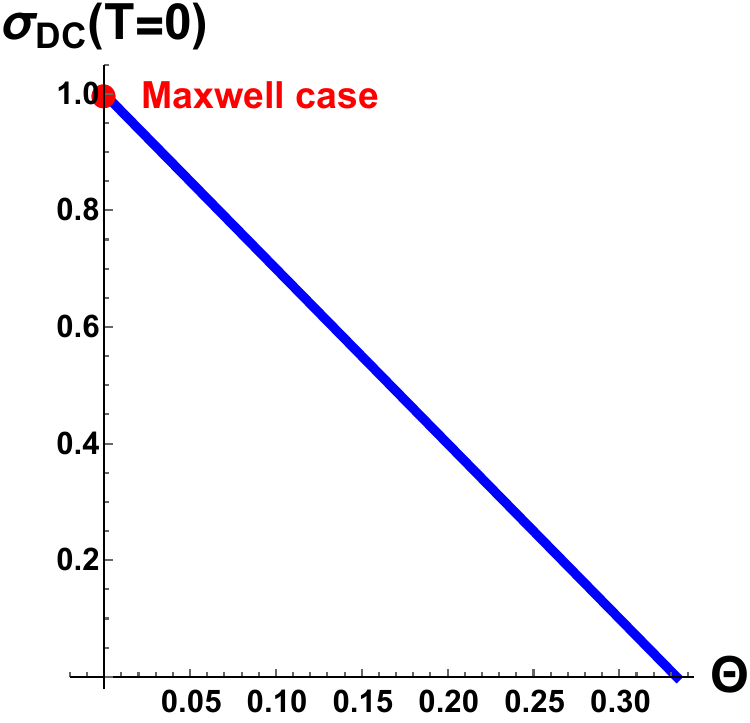}%
\qquad
\includegraphics[width=6cm]{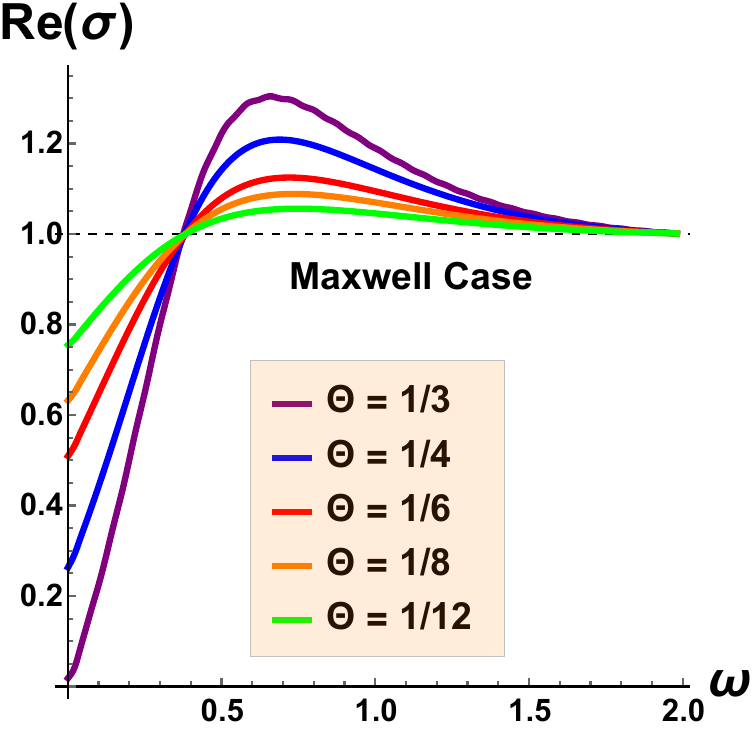}
\caption{iDBI \eqref{models} in the probe limit with zero momentum dissipation $\alpha=0$. We fix $\rho=1$. \textbf{Left:} DC conductivity at zero temperature in function of $\Theta$. \textbf{Right:} Optical Conductivity at $T=0.003$ for various values of $\Theta$  between the maximum one 1/3 to the Maxwell case $\Theta=0$.}
\label{ProbeFig}
\end{figure}
This formula explicitly shows how for $\Theta \leq 1/3$ the DC value is always less than the linear Maxwell case, which tells us that the iDBI choice \eqref{models} somehow obstructs the charge carriers' mobility, leading to an insulating behaviour.

One also notices that in the limit where $\Theta$ is very close to the maximum value $1/3$ and the corresponding ``material'' is a very good insulator a peak in the optical conductivity appears and gets sharper, see Fig.~\ref{ProbeFig}. 
An important point to underline is that the optical conductivity (in the probe limit as in the rest of the paper) shows a soft-gapped behaviour, while real Mott-Insulators are defined by an hard-gapped one\footnote{It has been recently argued that in `many body localized systems' there could exist a power law  $\sigma(\omega)\propto \omega^{\beta}$ behaviour at low frequency with $1<\beta<2$ \cite{nohard}. It would be nice to check and study further this scaling in our class of models.}. Within the present holographic models it appears very difficult to produce such hard gap. To reproduce this feature, it seems that one needs to resort to a dilaton with nontrivial running.

{We then go to the `probe limit I', where the background is completely dictated by the momentum dissipation sector, $\alpha \neq0$.}
Doing so in order to have a well defined extremal limit for our black hole we need to satisfy the bounds explained before and summarized in Fig.~\ref{probesigmatot1}. We emphasize that from now on, wherever not explicitly said, we are using the choice \eqref{models} with $\Theta=1$.
\begin{figure}[htbp]
\centering
\includegraphics[width=5cm]{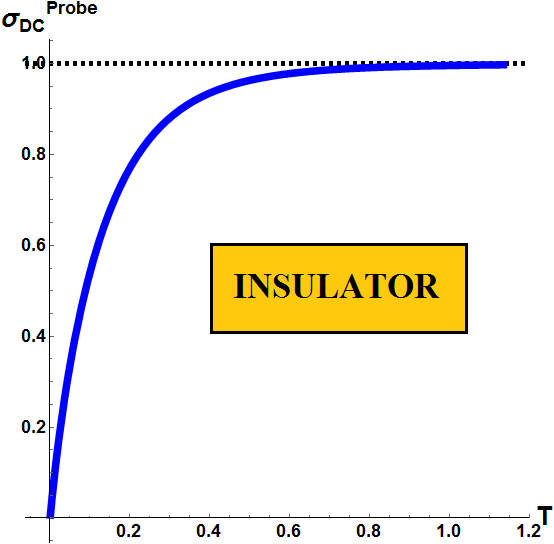}%
\qquad
\includegraphics[width=5cm]{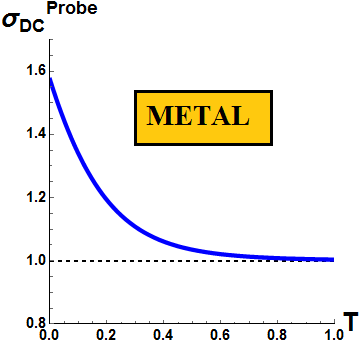}
\caption{Models \eqref{models} with $\Theta=\pm1$ in the probe limit with: $m^2=1.8,\,\alpha=\sqrt{2},\rho=1.8,V(z)=z$ ; \textbf{Left :} iDBI, showing an insulating behaviour; \textbf{Right :} DBI model exhibiting a metallic behaviour.}
\label{DCfigProbe}
\end{figure}

Figure \ref{DCfigProbe} shows that the non-linear extensions of Maxwell theory modify the DC conductivity (even in the probe vector limit), which now acquires a non trivial temperature dependence. At large temperature we recover the usual Maxwell (constant) result, as it should because large $T$ corresponds to small $F^2$ and we are considering only choices of $K(z)$ that go linear $K(z)\simeq z+\dots$ at small $z$. 
%
The low temperature behaviour, on the contrary, is very sensitive to the non-linearities and the particular model considered.

The iDBI \eqref{models} shows a strong suppresssion of the DC conductivity towards small temperatures and eventually a vanishing DC conductivity at $T=0$ can be obtained (by dialing $\Theta$ and/or $\mu$). This is the main reason to interpret these models as insulators. In a sense, in the  iDBI models the  electronic interactions are such that the charge carriers conductance is suppressed at small temperature.
Conversely, the usual DBI model increases the DC conductivity at small T exhibiting a typical metallic behaviour \cite{Karch:2007pd}. 

\begin{figure}[htbp]
\centering
\includegraphics[width=50mm]{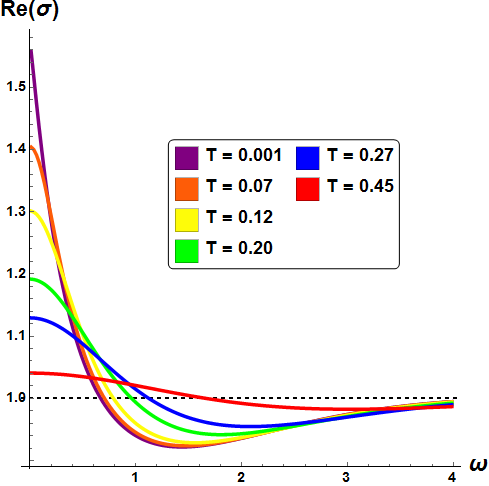}
\qquad\qquad
\includegraphics[width=50mm]{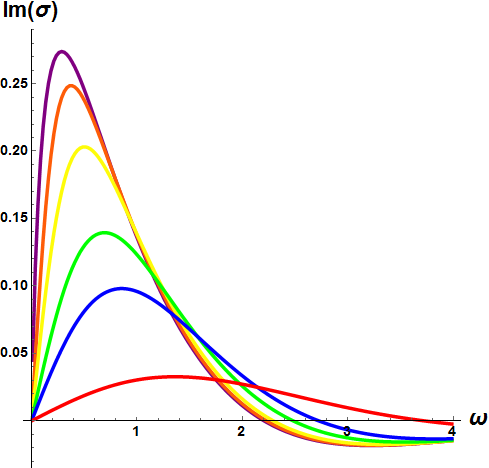}
\caption{AC conductivity for the DBI model (with $\Theta=-1$) in the probe vector limit with $m^2=1.8,\,\alpha=\sqrt{2},\rho=1.8,V(z)=z$ , the DC part is plotted in fig.\ref{DCfigProbe}. The metallic behaviour is evident.}
\label{DBIPPprobeAC1}
\end{figure}

\begin{figure}[htbp]
\centering
\includegraphics[width=50mm]{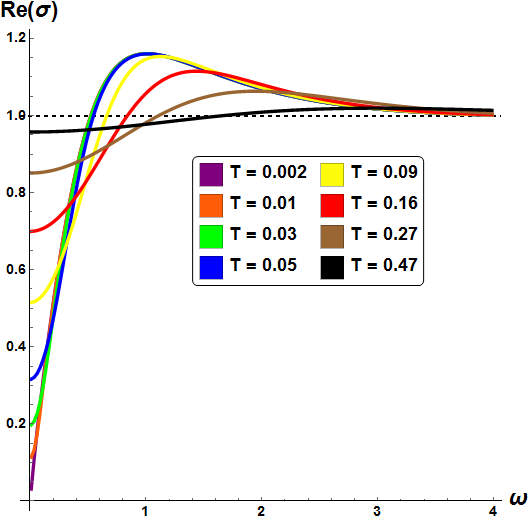}
\qquad\qquad
\includegraphics[width=50mm]{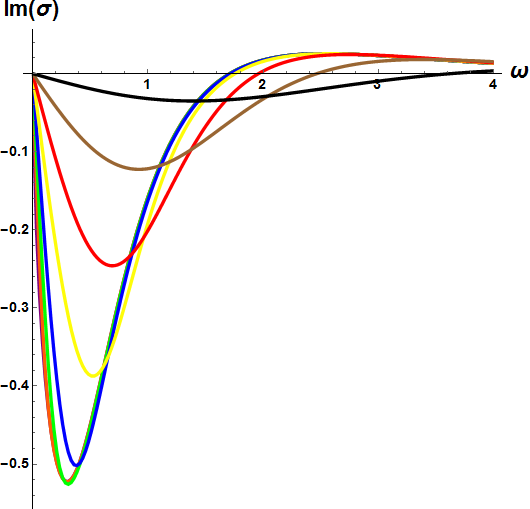}
\caption{AC conductivity for the iDBI (with $\Theta=1$) in the probe vector limit with $m^2=1.8,\,\alpha=\sqrt{2},\rho=1.8,V(z)=z$ , the DC part is plotted in fig.\ref{DCfigProbe}. The insulating behaviour appears clear. }
\label{DBImmprobeAC}
\end{figure}

We continue with the analysis of the optical conductivity for the choice \eqref{models}: the results are plotted in figures \ref{DBIPPprobeAC1} and \ref{DBImmprobeAC}.
In all the cases the large frequency limit coincides with the usual Maxwell conformal case (which in $2+1$ dimensions acquires the constant value $\sigma_\infty=1$); this is again a direct consequence of the fact that non linearities are relevant just in the IR region of the geometry while they dont affect any UV property.

The DBI model exhibits at small temperatures a clear Drude Peak behaviour which is characteristic of metallic materials; decreasing the temperature a smooth crossover drives into an incoherent metallic phase where the AC conductivity is almost constant and no dominating excitation appears. One can easily keep track of this dynamical behaviour studying the motion of the quasinormal modes of the system with temperature as done for a different setup in \cite{Davison:2014lua}.

The iDBI provides for a completely different phenomenology: there is a strong suppression of the conductivity at small frequencies and a spectral weight transfer to a `mid-infrared' peak. These features classify this model as an insulator. In this case the QNMs spectrum would be more similar to the one shown in \cite{Baggioli:2014roa}.\\
Note that in this section \eqref{probesection} the metric fluctuations are switched off and therefore we lack of a stress tensor in our CFT, this is the reason why we dont see any infinite in the DC conductivity and the reason why we do not need any dissipative mechanism to enter in the game. All the properties in this section just arise from the \textit{electron-electron} interactions provided by the non-linear electromagnetic extension (although the thermodynamics, at least in the second approximation, is also affected by the dissipative sector).

\subsection{Non-linear conductivity}

Let us now discuss the non-linear electrical response. The standard way to describe it is by exhibiting the nonlinear current-voltage ($J - E$) diagram, which encodes a nonlinear version of the electrical conductivity. We shall restrict to the DC case and to the simplest but nontrivial case to analyze, namely the probe vector limit where we neglect any mixing with the vector modes in the metric but still we keep all the nonlinear self-couplings of the gauge field. Similar studies have been done before, see \cite{Karch:2010kt,Sonner:2012if,Horowitz:2013mia}.

{We introduce an ansatz for the gauge field $A_t=A_t(u)$ and $A_x=E_x\,t + \delta A_x(u)$\footnote{Note that this ansatz differs from the one in \cite{Karch:2010kt,Sonner:2012if} in that we are including the `backreaction' of $\delta A_x(u)$ on $A_t$, which can be non-trivial starting at order $E^2$ for non-canonical kinetic terms. This suggests that at nonlinear level the $\mu-\rho$ curve also depends on $E^2$, but we shall not discuss this effect here.}. 
}
The NED-Maxwell equations in a black brane background with a generic emblackening factor $f(u)$ reduce to
\bea\label{NL}
&K'\left({-F^2\over2}\right)\,A_t' = -\rho \cr
&K'\left({-F^2\over2}\right)\,f(u)\,\delta A_x' = J_x \
\eea
where (since $K'(0)=1$) we already identified the integration constants $\rho$ and $J_x$ as the charge density and charge current. The field-strength invariant reduces to
\beq\label{F2}
z\equiv{-F^2\over2}=u^4 \left( A_t'^2 + {E_x^2 \over f(u)} - f(u) \delta A_x'^2 \right)
\eeq
Using \eqref{NL} and demanding $z$ to be regular at the horizon, one quickly obtains that
\beq\label{J-E}
J_x = K'\left({-F^2\over2}\right)\Big|_{u_h} \; E_x
\eeq
which nicely reproduces the linear conductivity result for small fields. 

However, \eqref{J-E} is much more informative now, as it holds for any value of $E_x$ and, indeed, $K'|_{u_h}$  depends nonlinearly on $E_x^2$. Let us make this dependence more manifest.

%
%
%

\begin{figure}[htbp]
\centering
\includegraphics[width=6cm]{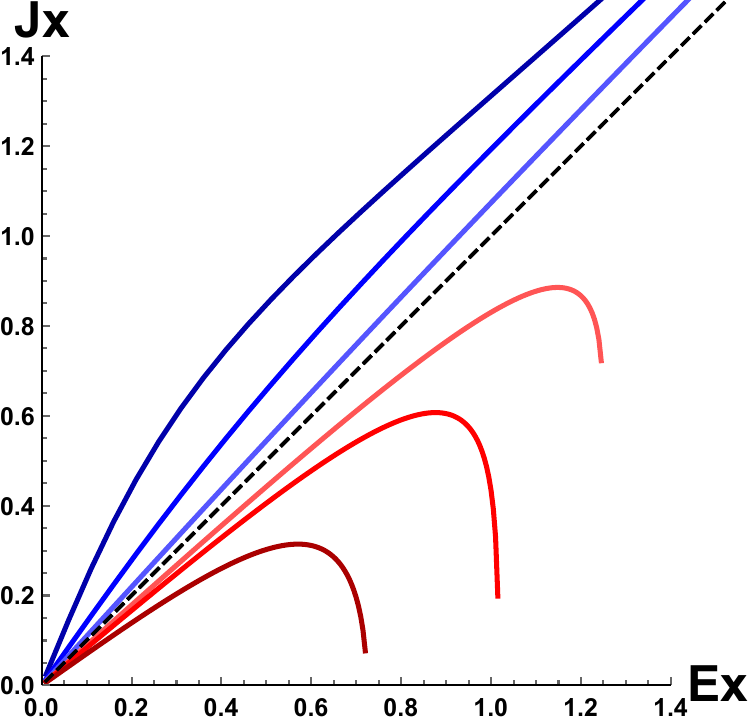}
\caption{Current-voltage ($J_x-E_x$) curve for the non linear models \ref{models} with parameters $\rho=u_h=1$ and $\Theta= -5,-1,-0.2,0,0.2,0.3,0.5$. Dashed line is the linear Maxwell theory. The background used for this plot is for simplicity taken to be RN but the qualitative results are insensitive to that.}
\label{NLsigma}
\end{figure}

To illustrate this, let us turn to the DBI/iDBI cases, the computation simplifies a lot. 
The equations of motion \eqref{NL} can be brought to the form
\beq\label{NL2}
-{A_t' \over \rho} ={ f(u) \delta A_x' \over  J_x } = \sqrt{\Theta \,E_x^2 \, u^4 + f(u) \over \Theta \,J_x^2\, u^4 +  (1-\Theta \rho^2 u^4)\,f(u)} 
\eeq
There is a choice of integration constants that maximizes the regularity of the solution, namely, that the argument of the square-root does not change sign anywhere. 
This could happen at the point $u_c=u_c(E_x^2)$ defined by 
\beq\label{uc}
{f(u_c)}+\Theta \,u_c^4\; E_x^2=0~.
\eeq
At the same point one must then require 
$$
J_x^2\, u_c^4 +  (1-\Theta \rho^2 u_c^4)\,f(u_c)=0
$$ 
so that 
the nonlinear conductivity is
\beq\label{sigmaNLiDBI}
\sigma_{NL} \equiv {J_x\over E_x}  = \sqrt{1- \Theta \,\rho^2\,u_c^4} ~.
\eeq
It is easy to check that this indeed coincides with $K'|_{u_h}$.

Note that for DBI models, $u_c<u_h$ whereas for iDBI models $u_c>u_h$. The IR regularity condition in that case originates from behind the horizon, which is certainly a bit intriguing. However, we do not find this to be necessarily a problem from any practical point of view (see Appendix \ref{NLgeneral} for more details about it).

We summarize in Fig.~\ref{NLsigma} the resulting nonlinear $J$-$E$ diagram. For DBI models, one clearly sees that at large $E$ one recovers $\sigma_{NL}=1+O(E^{-2})$. Therefore, even if the $J-E$ curve is entirely above the $J=E$ line, the nonlinear effect is to reduce the conductivity (from $\sqrt{1-\Theta \rho^2 u_h^4}$ to $1$). 
This is welcome since the Coulomb e-e interactions are expected if anything to reduce the conductivity. And as we see this does happen at high field even for the DBI case (which is the one where the linear conductivity is  enhanced with respect to the Maxwell case).

For the iDBI models the nonlinear effect is also to reduce the conductivity, but in a more dramatic way. Indeed, $\sigma_{NL} $ vanishes at some finite $E_x$. There is both a maximum current and voltage (or applied field $E_x$) that the material can withstand. 
Presumably, the region where the  $d J/d E < 0$ signals an instability, akin to the  electrical breakdown of insulators. The details of this instability are left for future investigation.

One also notices in Fig.~\ref{NLsigma} that the curves terminate at a maximal applied field $E_x$ in a branch-point fashion. This suggests that at that point one the gauge field becomes complex and therefore a naked singularity should re-appear. For such large applied voltage, then, one needs to do a proper analysis by including the backreaction. We defer this to future work.
\color{black}

Interestingly enough, nonlinear current-voltage curves quite close to  the ones for  the iDBI models have been measured in some lanthanide and cuprate compounds, \cite{NL0}. See also \cite{NL1,NL2} for other studies of the nonlinear electric response in strange metals.
\color{black}

\section{Discussion}\label{discussect}

Since we already give a summary of our results in Section \ref{sec1}, let us discuss now only the most salient features of the present results. The iDBI models \eqref{models} (with $\Theta>0)$, seem to behave effectively like Mott insulators in various different respects: i) the conductivity is unbounded from below, ii) there are MI transitions, iii) the conductivity can in some cases be even a decreasing function of carrier density, which appears like a clear smoking gun of traffic jam-like behavior. This happens in models where the only dynamical mechanism responsible for these features are the charge-sector self-interactions ({see \cite{Edalati:2010ww,Wu:2012fk,Edalati:2010ge,Ling:2014bda,Fujita:2014mqa,Ling:2015epa,Nishioka:2009zj,Kiritsis:2015oxa,Donos:2012js,Rangamani:2015hka,Donos:2014uba,Donos:2013eha} for more holographic models based on other mechanisms}). \color{black} Indeed, all of our models are scale invariant, they present no dynamical mass gap generation down to $T=0$, nor any  significant effects from the charge-disorder or charge-phonon couplings. In our view these already add up to quite significant evidence that these models do capture effectively Mott-like behavior.

We went on, then, to search for more characteristic features of these models, and we encountered 2 quite welcome `surprises'. First, we found that the models can support up to a maximum carrier density $\rho_\star$ which is set by the disorder strength (which can be thought of as the density of impurities) and the strength of the self-interactions. An upper bound on the carrier density like this (that holds only in the holographic models that display Mott-like behaviour) is quite reminiscent of the so-called `Mott criterium', that the carrier density must not exceed a certain bound (for 3 dimensional materials, $n^{1/3}\simeq a_0$ where $a_0$ is the Bohr radius \cite{Mott5,Ebeling:2008mg}. Obviously, the two bounds are qualitatively quite different, since the present one directly refers to disorder while the Mott criterium does not. However, we still find it quite striking and encouraging that there is an upper bound at all, given the disparity of the present approach compared to the traditional ones. In addition, having the upper bound depend on the disorder strength suggests that the basic mechanism that prevents the carrier concentration and which gives rise to the insulating behavior could be interpreted/related to Anderson-Mott localization.

The second surprise is that at densities close to (but below) this maximum density, our models generically develop an instability towards a superconducting state, implying that that the insulators that display Mott-like behavior should also exhibit an insulator-superconductor transition. This fits very well with the typical form of the High $T_c$  superconductor phase diagram (as in Fig.~\ref{SCsketch}) in function of the electron/hole doping. If confirmed, this has quite a few implications: even though the low-energy description does not unveil what is the underlying mechanism that drives this type of superconductivity (in terms of how the microscopic degrees of freedom `pair'), still it seems that one can link the mechanism behind the insulator-SC to the one operating in the metal-insulator transitions -- at least for those which are driven by electron-electron interactions  (and which are close to having a holographic dual). In particular, the value of the critical doping where the insulator-SC occurs should depend on disorder in a similar way as the maximal density $\rho_\star$ -- that is it should grow with disorder strength -- perhaps generically. Interestingly enough, this does happen in some materials \cite{disdome}.

We find that these facts support quite strongly that the  {\em effective holographic models} as developed in this work (allowing for quite general couplings in the Lagrangian), in the sense that they can be meaningfully used to model quite nontrivial phenomena in a clear, controlled and convenient way (the  ground state admits a  homogeneous description). 
%
Indeed, the effective holographic exercise makes good sense: there is a simple and clear map between the different actors in condensed matter (electrons, phonons, disorder) and the different ingredients in effective holography (the gauge field, the Stueckelberg  sector and whether it breaks translations mostly explicitly of spontaneously). Accordingly, there is a simple and clear correspondence with the kinds of interactions with different sectors. Interestingly, the various interactions between the gauge and Stueckelberg sectors match with what one expects from the condensed matter point of view. 
There are of course many points to develop and elaborate on, which we have to defer for future work.

In this connection, we have to comment on two more {\em (un)surprises}. The first one is the (likely) presence of `superluminal' modes in the iDBI model, that has also been found before in similar models \cite{Baggioli:2014roa,Kulaxizi:2015fza}. This certainly implies that the possible UV completion of the present models cannot be relativistic with the same notion of Lorentz invariance. However, this is by no means a problem for the effective holographic models.
Indeed, in the effective approach to the gauge/gravity duality, the starting point is that one takes the  gravity dual theory simply as a dynamical implementation of some strongly-coupled scale-invariant field theory. This can in principle be relativistic or not (this is just a matter of choice or convenience) and it is used to model the strongly coupled dynamics from the lowest energies to some `intermediate' cutoff scale below the ionic lattice spacing scale. For the gauge/gravity duality the case of  relativistic invariant CFTs is certainly the  best understood from a technical point of view, in part because it is the  {\em simplest} one to study -- it involves the least number of degrees of freedom/operators. From the condensed matter perspective, however, it certainly looks non-generic to assume that the scale invariant theory is relativistic. The only reason we assume this here so is by convenience -- these are the simplest models. It is clear, however, that this is meant to be interpreted as an emergent notion of Lorentz invariance, that is, with an emergent limiting speed that is of course constrained to be smaller than the speed of light. 
From this point of view, then, it is clear that the presence of modes that are faster than the {\em emergent} limiting speed is not at all a problem. On the contrary, it is a seemingly generic prediction that is perfectly compatible with observations because the emergent limiting speeds are normally subliminal -- as is the case for instance of graphene. In more realistic non-relativistic scale-invariant theories (and their possible holographic duals) the apparent issue raised by possible superluminal modes should not appear, but this certainly needs further investigation.

{The second (un)surprise  relates to the obvious concern that appears once one attempts to use holography as a low-energy effective description as we discuss here: how can a low-energy effective description (with effective Lagrangians that involve quite generic functions of operators) be predictive\footnote{See \cite{Khveshchenko:2016kmo} for a recent critique on the AdS-CMT panorama and its degree of success in this respect.\color{black}}? Fully answering this question is outside the scope of this article, but let us make one remark.}
The effective holographic Mott insulators constructed here crucially require large nonlinearities in the charge sector by way of non-canonical kinetic terms. It is quite reasonable then to expect enhanced effects also in the  {\em non-linear} electrical response (as was already pointed out in \cite{Baggioli:2014roa} for similar models). {We have initiated here a more detailed study of the nonlinear electrical response in the present models (previous studies were done in \cite{Karch:2010kt,Sonner:2012if,Horowitz:2013mia}) 
}
and found indeed very sharp effects in the nonlinear $\rho-\mu$ (susceptibility) curve as well as  in the nonlinear current-voltage ($J-E$). Mainly, the effects are in the form of a strong suppression of the susceptibility $d\rho/d\mu$ and conductivity $dJ/dE$ at  large density or voltage. 
These nonlinear effects can be {\em computed} in a controlled way ({\em i.e.}, within the regime of validity of the effective theory) and quite easily in our models and generalizations thereof.
Given that both the Mott-like insulating behavior (in the linear conductivity) and large nonlinear effects originate from the same gauge-field nonlinearities (a nonlinear $K(z)$), one would expect that the correlation between i) Mott-like behavior in the (linear) response and ii)  large nonlinear electrical response is a generic and robust. 
{Interestingly enough,  from the  experimental side the nonlinear regime can be accessed in some cases and large nonlinear effects along these lines have indeed been  observed \cite{NL0,NL1,NL2}.}
It is of course still not straightforward to extract how this linear-nonlinear correlation applies to real world materials -- in the first place because in many cases there can be more than one competing effect\footnote{
For instance, by having a departure from scale-invariance at $T=0$, which is modeled holographically with a running dilaton, see {\em e.g.} \cite{Donos:2013eha,Donos:2014uba,Gouteraux:2016wxj,Amoretti:2016cad}.
}. 
However, given that nonlinear transport/response includes a large set of observables, 
and given that the nonlinear response in the holographic models is basically fixed once the  linear response is fixed, we find that this an obvious class of phenomena to `test' these models, and there is even a potential to `explain' certain cross-correlations between linear- and nonlinear- response observables within these effective holographic descriptions\footnote{More work in this direction is underway \cite{nle}.}.

Admittedly, it is unclear whether this can be enough to make these models really useful. In any case, developing more the nonlinear response (electrical, thermal and of any sort) looks like a direction of potential great interest. We hope to return to this topic soon.

\section*{Acknowledgements}
We would like to thank Andrea Amoretti, Alessandro Braggio, Li Li, Wei-Jia Li, Yi Ling, Jaume Gazquez, Mikhail Goykhman, Saso Grozdanov, Elias Kiritsis, Blaise Gouteraux, Alexander Krikun, Nicodemo Magnoli, Daniele Musso,  Koenraad Schalm, Napat Poovuttikul, Philip Phillips, Sebastien Renaux-Petel, Stephane Roche and Ke Yang for useful discussions and suggestions. We acknowledge support from MINECO under grant FPA2014-55613-P,  DURSI under grant 2014SGR1450 and Centro de Excelencia Severo Ochoa program, grant SEV-2012-0234. MB is supported by a PIF grant from Universitat Autonoma de Barcelona UAB.
\appendix
\section{Consistency}\label{app1}
In this appendix we describe the minimal consistency conditions to ensure that our model stays healthy and free of patologies in both the vector and scalar (non-linear) sectors. We focus our analysis to the decoupling limit; of course a full treatment would be great but beyond the scope of this paper.\\
To achieve control on the correct renormalization of the various kinetic terms we use the trick of adding an external source to the second order lagrangian which reads:
\begin{equation}
\mathcal{\tilde{L}}^{(2)}\,\rightarrow\,\mathcal{L}^{(2)}\,-\,J_{ext}^\mu \,A_\mu\,\,\,\,\,\,,J_\mu^{ext}=\left(J_t,J_x,J_y,J_u\right)~.
\end{equation}
We then proceed in the decoupling limit (metric perturbations frozen) and analyze both the vector and the scalar modes contained in $A_\mu$. 
It is quite straightforward to check that the equation for the transverse-vector modes (which satisfy $\partial_i \delta A^i_v=0$) $\delta A_v^i$ of the perturbations  becomes:
\begin{equation}
\left[\partial^2_u+\left(\frac{f'(u)}{f(u)}-\frac{\PPi '(u)}{\PPi (u)} +\frac{2}{u}\right)\partial_u+\frac{\partial^2_y}{f(u)}-\frac{\partial^2_t}{f(u)^2}\right]\,\delta A_i\,=\,\frac{J_i}{2 \,u^2 f(u) \,K'\left(\PPi^2(u)\right)}\,.
\end{equation}
Requiring no ghosts in this sector  already gives a constraint on the function $K$, which reads
\begin{equation}
\boxed{\textit{No ghosts in the vector sector}\,\,\rightarrow\,\,K'(z)>0}
\end{equation}
Let's switch to the scalar degrees of freedom of the perturbations which are encoded in:
\begin{equation}
\delta A_{\mu}=\left(\delta A_t,\partial_x \zeta,\partial_y \zeta,\delta A_u\right)
\end{equation}
We use the gauge freedom to fix $\delta A_t=0$; we are left with three  non-independent equations,
\begin{align*}
&\frac{f}{K'(\PPi^2)}\,\partial_u\left(\frac{2\,\rho\,u}{\PPi'}\,\delta A_u\right)\,+\,\nabla^2\,\zeta\,=\,0\,,\\
&\frac{\PPi}{u^2\,f}\,\partial_u\left(\frac{u^2\,f}{\PPi}\,(\delta A_u\,-\,\partial_u \zeta)\right)\,+\,\partial^2_t\,\zeta\,=\,0\,,\\
&\nabla^2 \left(\delta A_u\,-\,\partial_u \zeta\right)-\frac{2\,\PPi}{u\,f\,\PPi'}\partial^2_t\,\delta\,A_u\,=\,\frac{J_u}{2 \,u^2\, K'\left(\PPi (u)^2\right)}\,.
\end{align*}
The easiest way to get a single equation for the single dynamical scalar mode (which we can identify as $A_u$) is to take the u-derivative of the first equation and substite it on the last one. We obtain:
\begin{equation}
\nabla^2 \delta A_u\,+\,\partial_u\left(\frac{f}{K'(\PPi^2)}\partial_u\left(\frac{2\,\PPi}{u\,\PPi'}\delta A_u\right)\right)-\frac{2\,\PPi}{u\,f\,\PPi'}\,\partial^2_t \delta A_u\,=\,\frac{J_u}{2\, u^2\, K'\left(\PPi (u)^2\right)}\,.
\label{scalpert}
\end{equation}
No ghosts in the longitudinal sector does not impose any further constraints on $K$ (note that $K'(z)>0$ implies $\PPi>0$).
\\
From eq.\ref{scalpert} we can read off the value of the `local' speed of sound (along the $x^i$ directions) for longitudinal modes, which reads:
\begin{equation}
c_s^2\,=\,\frac{u\,\PPi'}{2\,\PPi}
\label{speed}
\end{equation}
A conservative  requirement in order not to have gradient instabilities  in the longitudinal sector is that $c_s^2>0$, which leads to
\begin{equation}
\boxed{\textit{No gradient instabilities in the scalar sector}\,\,\,\rightarrow\,\,\,\PPi'(u)>0}
\end{equation}
\\
Note that using the Maxwell equation for the background metric \eqref{eoms} this condition can be rewritten into:
\begin{equation}
K'(z)+\,2\,K''(z)>0\,\,\,\rightarrow \,\,\,\left(\sqrt{z}\,K'(z)\right)'>0\,.
\end{equation}
All in all we get the following constraints:
\begin{align*}
\boxed{\, K'(z)>0\,, \,\,\,\left(\sqrt{z}\,K'(z)\right)'>0\,}
\end{align*}
where ${}'$ denotes always the derivative with respect to the argument of the function.\\
These can be rewritten in term of the $\PPi$ function appearing in the ansatz as:
\begin{align*}
\boxed{\, \PPi(u)>0\,, \,\,\,\PPi'(u)>0\,}
\end{align*}
Therefore requiring $c_s^2>0$ is equivalent to requiring that the canonical momentum $\PPi$ is a monotonous function of $u$.

In conclusion, from the expression \ref{speed} we can analyze the various situations which can appear in our model. The condition of having the speed of sound exactly unitary (in units of $c^2$) boils down to the condition of having the action of the Maxwell form and therefore $\PPi(u)\propto u^2$. The speed of sound can be rewritten in term of the K function as:
\begin{equation}
c_s^2\,=\,\frac{K'(\PPi^2)}{K'(\PPi^2)+2 K''(\PPi^2)}
\end{equation}
and if we compute it for the benchmark models with $p=1/2$ defined in \ref{models} we get:
\begin{equation}
c_s^2\,=\,\frac{1}{1-u^4\,\rho^2\,\Theta}
\end{equation}
which means that DBI model, corresponding to a metallic CFT, has  \textit{subluminal}  mode while the iDBI, representing an insulating CFT, shows potentially \textit{superluminal}  modes. This pattern is in accordance with what is known to happen in models with  non-canonical kinetic terms  \cite{Adams:2006sv}. This seems to be a common feature also (see  \cite{Kulaxizi:2015fza}) for other holographic models that to mimic the Mott physics through dipole fermions coupling in the bulk \cite{Edalati:2010ww,Edalati:2010ge}. 

Let us emphasize here that despite being certainly an issue to explore further, the possible problematic consequences of having  $c_s>1$ near the horizon are not quite clear in the context of holography. $c_s(u)$ does not represent the velocity of propagation of any particular mode or particle, so this does not directly conflict with the Lorentz symmetry of the underlying theory. Furthermore, in the low energy effective  interpretation of holography,  the velocity that we are implicitly setting to 1 does not necessarily correspond to the speed of light but perhaps the velocity of a light cone that emerges at low energies. From this low-energy perspective, the UV-completion of these scale invariant theories is actually expected (at least in the first step) completely non-relativistic  since it is given in terms of the atoms that build up the lattice.  For these reasons, we do not consider that $c_s>1$ jeopardizes the consistency of these setups. Let us remind the reader that exactly the same situation arises (in the phonon sector) for the models that give rise to insulating behavior due to electron-phonon interactions \cite{Baggioli:2014roa}.
\color{black}\\

Note that also in the Goldstones' sector we need to impose consistency constraints to avoid ghosts, gradient instabilities and to preserve Anti De Sitter as the UV asymptotical geometry.
Namely we have to ensure that:
\begin{align*}
\boxed{\,V'(X)\,>0\,,\,\,\,\,c_S^2\,=\,1\,+\,\frac{X\,V''(X)}{V'(X)}\,>0\,,\,\,\,V(0)\,=\,0\,}
\end{align*}
We refer to \cite{Baggioli:2014roa} for further details about the consistency issues of the model.\\

In the main text we consider models which preserve all these minimal conditions; it would be indeed interesting to perform a full consistency analysis which goes beyond the decoupling limit, however we do not expect major differences to appear.

\section{Conductivity}\label{app2}
In order to study the transport properties of the dual CFT we follow the conventional procedure \cite{Hartnoll:2009sz}. We are in particular interested in the electric response of the system\footnote{It would be really interesting to extend the study of the transport properties including thermal and magnetic responses; we leave it for future work.}. We switch on the perturbations defined as:
\begin{align*}
&\delta g_{ti}(t,u,y)\,=\,h_{ti}(t,u,y)\,,\\
&\delta g_{ui}(t,u,y)\,=\,\,h_{ui}(t,u,y)\,,\\
&\delta g_{ij}(t,u,y)\,=\,\frac{1}{u^2}\frac{\partial b(t,u,y)}{\partial y}\, ,\\
&\delta A_i(t,u,y)\,=\,a_i(t,u,y)\, ,\\
&\delta \phi(t,u,y)\,=\,\Phi(t,u,y)\,.
\end{align*}
We then proceed linearizing the equations of motion for those perturbations in a gauge-invariant picture.
Aside from $a_i$, we use the following gauge-invariant combinations:
\beq
T_i \equiv u^2 h_{ti} - {\partial_t \Phi_{i} \over \alpha} \; , \;
U_{i}  \equiv  f(u)\big[h_{ui} -  {\partial_u \Phi_{i}\over \alpha u^2}\big] \; , \;
B_i \equiv   b_i -{\Phi_i\over\alpha} ~.\nonumber
\eeq
We use also the following definition for the Fourier decomposition of all the fields:
\begin{equation}
\zeta^A(u,t,y)\,=\,\,e^{-i \,(t \,\omega\, -\,k\, y)}\,\zeta^A(u)
\label{fourierexp}
\end{equation}
With these choices we are left with the following equations (see \cite{Kodama:2003kk} for the generic structure and procedure) :
\begin{align}
&\scriptstyle{-2 \,q^2\,\rho\, u^2 \,a'-\frac{i \,k^2\, \omega B}{f}-\frac{i\, u^2\, \omega\,  U f'}{f^2}+\frac{i\, u^2\, \omega \, U'}{f}-T \left(\frac{2\,
   \alpha^2 \,m^2 V'\left(\alpha^2 u^2\right)}{f}+\frac{k^2}{f}\right)-\frac{2
  T'}{u}+T''\,=\,0}\nonumber\\
&\scriptstyle{k\left[B''+B' \left(\frac{ f'}{f}-\frac{2 }{u}\right)+B \left(\frac{ \omega ^2}{f^2}-\frac{2\, \alpha^2 
   m^2 V'\left(\alpha ^2 u^2\right)}{f}\right)-\frac{i \,\omega\,  T}{f^2}-\frac{ u^2 U'}{f}\right]\,=\,0\,,}\nonumber\\
   &\scriptstyle{i \,q^2\,\rho\,  u^5\, \omega\,  a\, f+\frac{1}{2} k^2 u^3 f^2 B'+U \left(-\alpha^2\, m^2\,u^5\, f\, V'\left(\alpha ^2
   u^2\right)-\frac{1}{2} k^2 u^5 f+\frac{u^5 \omega ^2}{2}\right)-\frac{1}{2} i \,u^3\, \omega\,  f\, T'\,=\,0\,,}\nonumber\\
   &\scriptstyle{k^2\,B\, V'\left(\alpha ^2 u^2\right)-\frac{i \,\omega  \,T\, V'\left(\alpha ^2 u^2\right)}{f}-2 \,\alpha ^2 \,u^3 \,U\,
   V''\left(\alpha ^2 u^2\right)-u^2 \,U'\, V'\left(\alpha ^2 u^2\right)\,=\,0\,,}\nonumber\\
   &\scriptstyle{a''+a' \left(\frac{f'}{f}-\frac{\PPi '}{\PPi}+\frac{2}{u}\right)+\frac{a \left(\omega ^2-k^2
   \,f\right)}{f^2}-\frac{\PPi\, T'}{u^2 f}-\frac{i \,\omega \, \PPi \, U}{f^2}\,=\,0\,.}
   \label{ComplEQ}
\end{align}
where we kept implicit all the u dependences.\\
The first four equations displayed in \eqref{ComplEQ} are not indipendent and we can forget about the first one. 
In addition it is easy to see that $T(u)$ is completly constrained and reads:
\begin{equation}
T(u)\,=\,\frac{i f(u) \left[k^2 B(u)+u^2 \left(-\frac{2 \,\alpha ^2\, u \,U(u) \,V''\left(\alpha ^2 u^2\right)}{V'\left(\alpha ^2
   u^2\right)}-U'(u)\right)\right]}{\omega }\,.
\end{equation}
We can therefore eliminate the gauge invariant combination T in favor of the others one. At finite momentum k we are left with the system of coupled equations given by:
\begin{eqnarray*}
&\scriptstyle{\frac{\PPi(u)}{u^2}\p_u\left(\frac{u^2}{\PPi(u)}\,f(u)\, a'(u)\right)+\left(\frac{\omega^2}{f(u)}-k^2
-2\,q^2\,\rho\,\PPi(u)\right)a(u)\,=\,+\frac{i\,\PPi(u)\,(2 M^2(u)+k^2)}{\omega}U(u)-\frac{i \,f(u)\, \PPi(u) \,k^2}{u^2 \,\omega}B'(u)}\\
&\scriptstyle{\frac{1}{u^2}\p_u\left[\frac{f(u)\,u^2}{M^2(u)}\p_u(M^2(u)U(u))\right]+\left(\frac{\omega^2}{f(u)}-k^2-2\,M^2(u)\right)U(u)\,=\,-2\,i\,q^2\,\rho\,\omega\,a(u)+\frac{f'(u)\,k^2}{\omega}B(u)}\\
&\scriptstyle{k\left[u^2\,\p_u\left(\frac{f(u)}{u^2}B'(u)\right)+\left(\frac{\omega^2}{f(u)}-k^2-2\,M^2(u)\right)B(u)\,=\,-2 \frac{M'(u)}{M(u)}U(u)\right]}
\end{eqnarray*}
where $M^2(u)=\alpha^2\,m^2\,V'(u^2\alpha^2)$.\\
It is straightforward to check that in the linear case for the Maxwellian sector $\PPi(u)=\rho \,u^2$ (and setting $q=1$) we recover exactly the equations of \cite{Baggioli:2014roa}.\\
In the homogeneous case ($k=0$) (and $q=1$) we can simplify the problem (see \cite{Andrade:2013gsa} for details about the trick) to a
2X2 system of equations which reads:
\begin{align}
&\frac{1}{u^2}\p_u\left[\frac{f(u)\,u^2}{M^2(u)} v'(u)\right]+\frac{\omega^2}{f(u)\,M^2(u)}v(u)-2\,v(u)-2\,\rho\,a(u)=0\\
&\frac{1}{u^2}\p_u\left[\frac{f(u)\,u^2}{\PPi(u)} a'(u)\right]+\frac{\omega^2}{f(u)\,\PPi(u)}a(u)-2\,v(u)-2\,\rho\,a(u)=0
\label{perteqns}
\end{align}
It is straightforward to check that the mass matrix has zero determinant meaning that there is a massless mode which permits to run the usual argument to get the analytical formula for the DC conductivity \cite{Blake:2013bqa}.\\
We will derive the DC conductivity using the elegant and simple method described in \cite{Donos:2014cya} in a gauge invariant formalism.\\
For the sake of computing the DC conductivity, we consider a different ansatz with respect to \eqref{fourierexp}. In particular we consider homogeneous modes ($k=0$) and a constant electric field in the x direction, which correspond to the configuration given by:
\begin{align}
&T(t,u,y)\,=\,T(u)\,,\nonumber\\
&U(t,u,y)\,=\,U(u)\,,\nonumber\\
&B(t,u,y)\,=\,B(u)\,,\nonumber\\
&a(t,u,y)\,=\,-E_x \,t +a(u)\,.
\end{align}
Because of the homogeneous choice we are left with four equations which dont involve the B field and which read:
\begin{align}
&U(u)-\frac{E_x\,q^2\, \rho }{\alpha ^2 \,m^2\, V'\left(\alpha ^2 u^2\right)}\,=\,0\,,\nonumber\\
&\frac{2 \,\alpha ^2 \,u \,U(u)\, V''\left(\alpha ^2 u^2\right)}{V'\left(\alpha ^2 u^2\right)}+U'(u)\,=\,0\,,\nonumber\\
&-2 \,q^2\,\rho\,  u^2\, a'(u)-\frac{2\, \alpha ^2 \,m^2\, T(u) V'\left(\alpha ^2 u^2\right)}{f(u)}-\frac{2\, T'(u)}{u}+T''(u)\,=\,0\,,\nonumber\\
&a''(u)+a'(u) \left(\frac{f'(u)}{f(u)}-\frac{\PPi '(u)}{\PPi (u)}+\frac{2}{u}\right)-\frac{\PPi (u) T'(u)}{u^2 f(u)}\,=\,0\,.
\label{DCeq}
\end{align}
The first two equations consistently imply:
\begin{equation}
U(u)\,=\,\frac{E_x\,q^2\, \rho }{\alpha ^2 \,m^2\, V'\left(\alpha ^2 u^2\right)}\,.
\label{refU}
\end{equation}
Maxwell equations reads, as expected, as the radial conservation of a quantity:
\begin{equation}
\partial_u\left(-\frac{u^2\,\rho\,f(u)}{\PPi(u)}a'(u)+\,\rho\,T(u)\right)\,=\,0\,.
\end{equation}
which is going to correspond to the electric current J of the dual field theory:
\begin{equation}
J_x\,=\,-f(u)\,K'(\PPi(u)^2)\,a'(u)+\,\rho\,T'(u)\,.
\label{current}
\end{equation}
This current can be computed at any value of the radial coordinate including the horizon position $u=u_h$.\\
The DC conductivity can be then computed dividing the expression \eqref{current} by the constant electric field in the x direction:
\begin{equation}
\sigma_{xx}\,=\,\frac{J_x}{E_x}
\end{equation}
In order to do so we need to compute the current J in terms of the horizon data and in particular we have to find the horizon behaviour of the fluctuations.\\
The key point is to impose the regularity at the horizon for the U(1) gauge field ($F^2$ with finite norm at the horizon); at the level of the perturbation $F^2$ reads:
\begin{align}
F^2\,=\,&2 \,u^4 f(u) a'(u)^2-4 u^2 T(u) \PPi (u) a'(u)-\frac{2 E_x^2 u^4}{f(u)}-\frac{4E_x u^4 \PPi (u)
   U(u)}{f(u)}\,+\nonumber\\
   &+\frac{2 T(u)^2 \PPi (u)^2}{f(u)}-\frac{2 u^4 \PPi (u)^2 U(u)^2}{f(u)}
\end{align}
Demanding that flux to be finite we need to impose the following regularity conditions at the horizon:
\begin{equation}
a'(u)\,=\,-\frac{E_x}{f(u)}\,\,,\,\,\,\,\,\,\,\,\,T(u)\,=\,u^2\,U(u)
\end{equation}
where $U(u)$ is given by \eqref{refU}.\\
All in all the current $J_x$ computed at the horizon position reads:
\begin{equation}
J_x\,=\,E_x\, K'\left(\PPi (u_h)^2\right)+\,\frac{E_x\,q^2\, \rho^2\, u_h^2 }{\alpha^2 m^2 V'\left(\alpha ^2 u^2\right)}
\end{equation}
We can finally proceed to define the DC electric conductivity just in terms of horizon data .\\
The final formula for the DC conductivity is a generalization of the one got in (\cite{Baggioli:2014roa}) and in terms of the horizon quantities reads:
\begin{equation}
\sigma_{DC}=K'(\PPi^2(u_h))+\frac{q^2\,\rho^2 \,u_h^2}{M^2(u_h)}
\end{equation}
where $M^2(u_h)\,=\,\alpha^2 m^2 V'\left(\alpha ^2 u_h^2\right)$ .

\section{Non-linear conductivity for general $K(z)$}\label{NLgeneral}

Using \eqref{NL} in \eqref{F2}, and enforcing the regularity at the horizon by \eqref{J-E}, one arrives at the following relation valid for any choice of $K(z)$,
\beq\label{Kp1}
z K'(z)^2=u^4 \left( \rho^2 + E_x^2 \;{K'(z)^2-K'(z_h)^2\over f(u)}  \right)~,
\eeq
with  $z(u)={-F^2/2}$ on the solution.
This equation is manifestly consistent with regularity of $z$ at the horizon. From this equation one can in principle extract how $z$ depends on $u$ and on $E_x$.
It is clear that for the linear case ($K'=1$), $z=\rho^2 u^4$ --  in particular it does not depend on $E_x$; but in the general case, $z$ (and therefore $K'(z_h)$ does depend on $E_x$. The way to fix $K'(z_h)$ then requires an additional boundary condition.
In non-canonical theories like this, there are additional regularity conditions that arise naturally which end up fixing $z_h$.

%
%

One can recast \eqref{Kp1} as
\beq\label{Kp21}
K'(z)^2= u^4 \;{ E_x^2 K'(z_h)^2 -\rho^2 f(u) \over E_x^2 u^4 - z\; f(u)}
\eeq
There is therefore a natural boundary condition that maximizes the regularity of the solution, namely that the r.h.s of \eqref{Kp21}  does not change sign anywhere. This could happen at the point $u_{\tilde c}$ defined by 
\beq\label{uctilde}
E_x^2 u_{\tilde c}^4 - z(u_{\tilde c})\; f(u_{\tilde c})=0~.
\eeq
The natural regularity condition then demands that at $u_{\tilde c}$ the numerator of \eqref{Kp21} also vanishes there,
that is, 
$$
K'(z_h)^2 ={\rho^2 f(u_{\tilde c}) \over  E_x^2 } = {\rho^2 u_{\tilde c}^4 \over  z(u_{\tilde c})} 
$$
Since $z(u)$ is a smooth function of $u$, \eqref{uctilde} shows that  $u_{\tilde c}$ depends on $E_x$ smoothly, with $u_{\tilde c}=u_h+{\cal O}\big(E_x^2\big)$ for small $E_x$. 

Therefore the nonlinear DC conductivity ${J\over E}=K'(z_h)$ depends smoothly on $E_x^2$ in general, possibly up to some maximum value of $E_x$ where the roots of \eqref{uctilde} cease to exist.
Let us emphasize that the radius $u_{\tilde c}$ is not the same as the $u_c$ introduced in \eqref{uc}. Still, the result for the nonlinear conductivity (namely how $K(z_h)$ depends on $E_x$) does not depend on whether one uses the regularity condition at $u_c$ or at $u_{\tilde c}$

\end{document}